\documentclass[twocolumn, trackchanges]{aastex701}

\usepackage{subcaption}
\usepackage{tabularx} 
\usepackage{booktabs}
\usepackage{longtable}
\usepackage{placeins}
\usepackage{amsmath,amssymb, amsthm, amsfonts}

\begin{document}

\title{Blazar classification from multi-wavelength data using Artificial Neural Network}

\author[orcid=/0009-0009-6010-8682,gname=Saqlain, sname='Afroz']{Saqlain Afroz}
\affiliation{Indian Institute of Science Education and Research Kolkata-741246, India}
\email{sqn3680@gmail.com}  

\author[orcid=0009-0004-0580-8823,gname=Titir, sname='Mukherjee']{Titir Mukherjee} 

\affiliation{Clemson University, Clemson-29634, U.S.A.}
\email{tmukher@clemson.edu}

\author[orcid=0000-0002-1173-7310,gname=Raj,sname=Prince]{Raj Prince}
\affiliation{Department of Physics, Institute of Science, Banaras Hindu University, Varanasi-221005, India}
\email{priraj@bhu.ac.in}

\correspondingauthor{Raj Prince} 
\email{sqn3680@gmail.com, priraj@bhu.ac.in}


\begin{abstract}
The Fermi Large Area Telescope (Fermi-LAT) has detected more than 7,000 gamma-ray sources, a significant fraction of which are identified as blazars, while a comparable number remain classified as blazars of uncertain type (BCUs) or are unassociated with counterparts at other wavelengths. The absence of complete multi-wavelength spectral information presents a major obstacle to robust source classification, despite such data providing the most reliable means of understanding blazar properties. In this work, we focus on classifying BCUs into the two primary blazar subclasses, flat-spectrum radio quasars (FSRQs) and BL Lacertae objects (BL Lacs), using a feed-forward artificial neural network (ANN) trained on multi-wavelength observational parameters. We first identify the most informative features by quantifying their information content and then use these features to train the ANN, whose performance is evaluated using a $k$-fold cross-validation strategy to ensure robust generalization. The trained model is subsequently applied to classify BCUs into BL Lacs and FSRQs. Our results demonstrate that machine learning–based classification using a carefully selected set of multi-wavelength parameters offers an efficient and reliable approach for resolving the nature of BCUs and improving the completeness of the gamma-ray blazar population in Fermi-LAT catalogs.

\end{abstract}


\section{Introduction}

Blazars constitute a radio-loud subset of Active Galactic Nuclei (AGNs) and represent some of the most luminous extragalactic phenomena observed in the $\gamma$-ray sky. Characterized by relativistic jets oriented close to the observer's line of sight, their Spectral Energy Distribution (SED) typically exhibits a bimodal structure in the $\log{\nu} - \log{\nu F_{\nu}}$ plane, spanning the electromagnetic spectrum from radio to $\gamma$-rays. The emission is dominated by non-thermal processes: the low-energy component arises from synchrotron radiation produced by relativistic electrons within the jet's magnetic field, while the high-energy component is generally attributed to Inverse Compton (IC) scattering of low-energy photons by the same electron population.

Conventionally, blazars are classified based on the characteristics of their optical spectra. Flat Spectrum Radio Quasars (FSRQs) display strong, broad emission lines, whereas BL Lacertae objects (BL Lacs) exhibit featureless spectra or very weak emission lines. Alternatively, classification can be based on the broadband SED, specifically the location of the synchrotron peak and total jet power \citep{1998MNRAS.299..433F}. Following the launch of the \textit{Fermi} Large Area Telescope (LAT), the population of detected $\gamma$-ray blazars has expanded significantly. However, a substantial fraction of these sources lack definitive classification into the established BL Lac or FSRQ subtypes. These objects are designated as Blazar Candidates of Uncertain type (BCUs).

The primary challenge in classifying BCUs lies in the scarcity of multi-wavelength data. Current all-sky monitoring telescopes often suffer from limited source localization accuracy (e.g., several arcminutes for \textit{Fermi}-LAT), leading to positional ambiguity and multiple potential counterparts. This hinders the construction of complete broadband SEDs and the acquisition of optical spectra necessary for spectroscopic classification. The proliferation of BCUs is non-trivial; since the release of the first \textit{Fermi} catalog (1FGL), the proportion of unclassified objects has risen from $14\%$ to $42\%$ in the fourth catalog (4FGL). Consequently, there is an urgent need for robust, automated algorithms capable of distinguishing between BL Lacs and FSRQs without relying solely on elusive optical spectroscopy. Accurate classification is crucial not only for understanding jet physics but also for constraining the Extragalactic Background Light (EBL) and optimizing target selection for future observatories such as the Cherenkov Telescope Array (CTA).

In recent years, machine learning (ML) has emerged as a powerful tool for blazar classification, with numerous studies demonstrating the efficacy of statistical models. \cite{Kang:2019cck} applied Support Vector Machines (SVM), Artificial Neural Networks (ANN), and Random Forests (RF) to classify 1,312 BCUs from the 4FGL catalog. Using 23 $\gamma$-ray parameters, they achieved an accuracy of 92.9\%, notably finding that increasing the feature set size did not necessarily enhance performance. Similarly, \cite{Kovacevic:2020sly} utilized ANNs trained on $\gamma$-ray light curves and spectral data to classify BCUs with 90\% precision, providing a viable method for preliminary analysis when optical data is unavailable.

To address uncertainties in classification, \cite{Butter:2021mwl} introduced Bayesian Neural Networks (BNNs), which offer probabilistic outputs alongside decision boundaries. Their work demonstrated that data augmentation could improve robustness in small, imbalanced datasets. Furthermore, \cite{Agarwal:2023vra} implemented an ensemble "unanimous voting" approach across five algorithms (including XGBoost and CatBoost), achieving an AUC score of 0.96 and improving prediction confidence. More recently, \cite{Bhatta:2023qrm} focused on model portability and minimal feature reliance using self-supervised learning, while \cite{Gharat:2024uof} expanded the scope by introducing a multi-task architecture for simultaneous classification and redshift estimation.

Despite these advancements, the majority of existing frameworks rely predominantly on $\gamma$-ray data \citep{Kang:2019cck, Kovacevic:2020sly}. While some efforts have been made to incorporate multi-wavelength properties \citep{2019ApJ...887...18K, 2020ApJS..248...23D}, there remains a significant opportunity to enhance classification reliability by systematically integrating broadband catalog-level features.

Motivated by this gap, this work presents a Deep Neural Network (DNN) approach that utilizes multi-wavelength data to categorize BCUs. By moving beyond $\gamma$-ray-only features, we aim to capture a more diverse set of physical characteristics, thereby providing a robust framework scalable to future high-volume astronomical catalogs.

The main contributions of this work are summarized as follows:

\begin{itemize}
    \item We develop a machine learning framework that leverages multi-wavelength catalog-level features to classify Blazar Candidates of Uncertain type (BCUs). This approach enables a data-driven characterization based on broadband observational properties, eliminating the reliance on exclusive, handcrafted classification criteria.
    \item We apply the proposed methodology to well-defined samples of BL Lac objects, FSRQs, and BCUs, demonstrating the model's effectiveness in robust subclass discrimination.
    \item We systematically evaluate model performance using a comprehensive suite of statistical metrics, thereby ensuring robustness and minimizing the impact of catalog-specific or sampling-related biases.
    \item We interpret the classification results through the lens of intrinsic variability behavior, highlighting the physical distinctions between blazar subclasses beyond purely statistical performance.
\end{itemize}

The remainder of this paper is organized as follows: Section \ref{data} outlines the data acquisition strategy, including the spatial cross-matching of datasets, and details our neural network architecture and training methodology. In Section \ref{result_and _disc}, we present and discuss the classification results. Finally, Section \ref{conclusion} summarizes our findings and concludes the paper.

\section{Data Selection and Model Training} \label{data}

To construct a robust framework for blazar classification, we integrated multi-wavelength observational data from three authoritative astronomical catalogs: the \textit{Fermi}-LAT Fourth Source Catalog (4FGL-DR4) \citep{Ballet:2023qzs}, the ROMA-BZCAT v5 Multi-frequency Catalog of Blazars \citep{Massaro:2015nia}, and the Third LAT AGN Catalog (3LAC) \citep{Fermi-LAT:2015bdd}. This synthesis provides comprehensive coverage across the radio, optical, X-ray, and $\gamma$-ray bands, establishing a diverse feature space for differentiating blazar subclasses. It is important to note that complete multi-wavelength information has rarely been used for classification tasks. Nevertheless, such data provide a more comprehensive view of blazars, making it essential to investigate how features across different wavelength bands encode information about the source and how this information can be leveraged to improve source classification.

\subsection{Catalog Description}
The 4FGL-DR4 catalog \citep{Ballet:2023qzs} constitutes the primary source of high-energy $\gamma$-ray data used in this work. It represents the latest data release of the Fourth \textit{Fermi}-LAT Source Catalog and is based on approximately 14 years of observations collected by the \textit{Fermi} Large Area Telescope, providing detailed spectral and variability information for a large population of active galactic nuclei (AGN). Ground-truth class labels required for supervised learning were obtained from the ROMA-BZCAT v5 catalog \citep{Massaro:2015nia}, which offers optically confirmed classifications of blazars, including BL Lacertae objects (BL Lacs) and Flat-Spectrum Radio Quasars (FSRQs). In addition, the 3LAC catalog \citep{Fermi-LAT:2015bdd} was used to supplement the dataset with well-established spectral properties of \textit{Fermi}-detected AGN, such as photon indices and synchrotron peak classifications. These parameters play a crucial role in characterizing the spectral energy distributions (SEDs) of blazars and provide complementary information relevant for classification tasks.

\subsection{Dataset Construction}
We carried out spatial cross-matching of source positions, specified by Right Ascension and Declination, using the TOPCAT software \citep{Taylor_2017}. A matching radius of 10 arcseconds was adopted to retain only high-confidence associations between the catalogs. This process resulted in the construction of three distinct datasets, which were used to systematically examine how different sets of features influence the classification performance. These datasets are:

\begin{itemize}
    \item \textbf{Dataset I:} This dataset was constructed by cross-matching the ROMA-BZCAT v5 catalog with the 4FGL-DR4 catalog, resulting in 1,765 matched sources. The feature set consists of standard multi-wavelength photometric and spectroscopic parameters: the radio flux at 1.4~GHz ($F_R$), the optical magnitude in the $R$ band ($R_{\mathrm{mag}}$), the X-ray flux in the 0.1-2.4~$keV$ energy range ($F_X$), and the redshift ($z$). After applying the filter, the final sample contains 1,499 sources, comprising 809 BL Lac objects and 690 FSRQs.

    \item \textbf{Dataset II:} This dataset was obtained by cross-matching the ROMA-BZCAT v5 catalog with the 3LAC catalog, yielding 1,142 matched sources. The feature space includes the redshift ($z$), the synchrotron peak frequency ($\nu_{\mathrm{peak}}$), the spectral index (derived assuming a log-parabola or the default spectral model), and the power-law index (assuming a simple power-law spectrum). Applying the corresponding selection criteria resulted in a final sample of 988 sources, consisting of 567 BL Lac objects and 421 FSRQs.

    \item \textbf{Dataset III:} This dataset was constructed using the same cross-matching procedure as Dataset~II, but with an extended feature set. In addition to the spectral parameters used in Dataset~II, namely the synchrotron peak frequency ($\nu_{\mathrm{peak}}$), the spectral index, and the power-law index, we include the multi-wavelength observables from Dataset~I: the radio flux at 1.4~GHz ($F_R$), the optical $R$-band magnitude ($R_{\mathrm{mag}}$), the X-ray flux in the 0.1-2.4~$keV$ range ($F_X$), and the redshift ($z$). This results in a total of seven input features. The final sample size remains 988 sources, with 567 BL Lac objects and 421 FSRQs.
\end{itemize}

These features were selected to span the full electromagnetic spectrum, capturing the distinct physical processes governing blazar emission and facilitating the discrimination of subclasses within a high-dimensional space.


\subsection{Neural Network Architecture and Training Strategy}

To classify blazar candidates using the multi-wavelength feature vectors described in Section~\ref{data}, we implemented a feed-forward Artificial Neural Network (ANN) within the PyTorch framework \citep{paszke2019pytorchimperativestylehighperformance}. The model follows a Multi-Layer Perceptron (MLP) design, chosen to balance representational power with model simplicity and to reduce the risk of overfitting given the available sample size.

\begin{figure}[htbp!]
    \centering
    \includegraphics[width=0.9\linewidth]{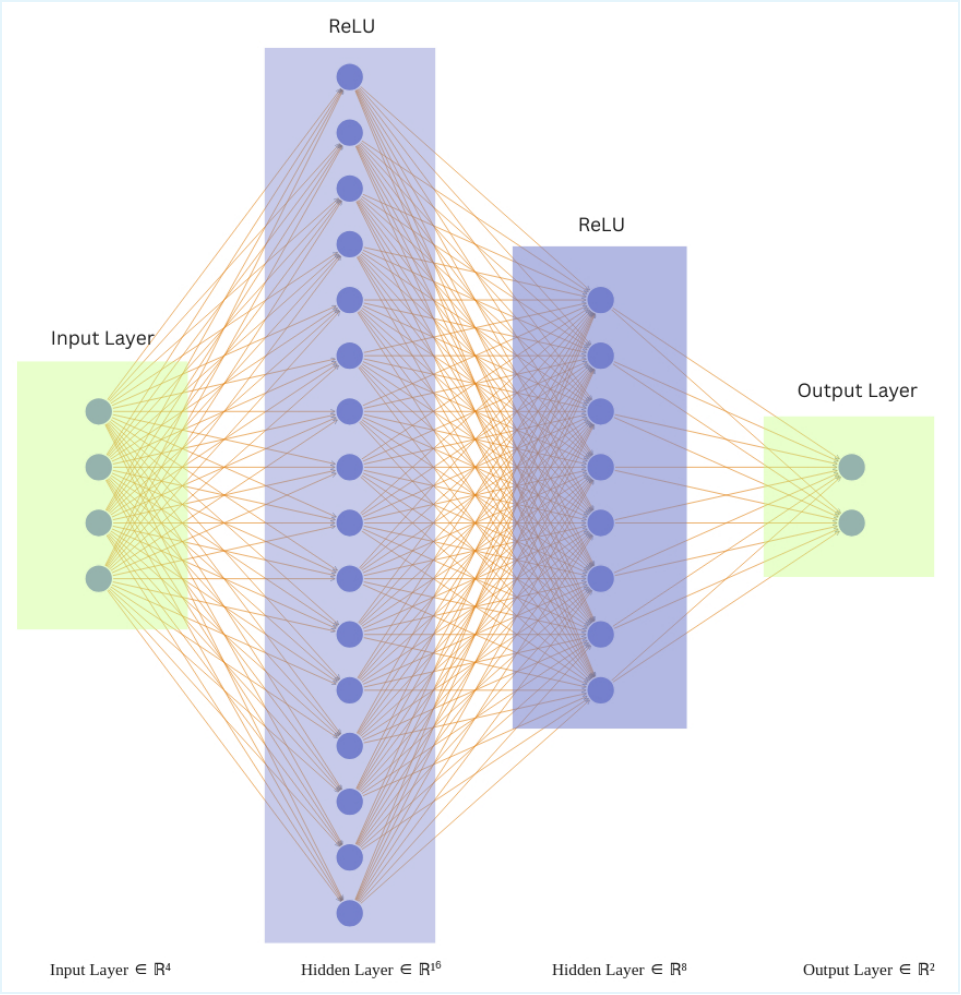}
    \caption{Schematic of the feed-forward Artificial Neural Network (ANN) used for binary classification. The architecture processes input vectors (dimension 4 or 7) through two hidden layers using ReLU activation functions. The final output layer provides class scores for BL Lac and FSRQ. Training is performed using the AdamW optimizer and cross-entropy loss.}
    \label{fig:fnn-arc}
\end{figure}

The network takes as input a feature vector of dimension $D_{in} \in \{4, 7\}$, corresponding to Dataset~I, Dataset~II, and Dataset~III, respectively. The architecture consists of a sequence of fully connected (dense) layers interleaved with non-linear activation functions. The input features are projected into a $16$-dimensional hidden space and passed through Rectified Linear Unit (ReLU) activation function, which is again passed through an $8$-dimensional hidden space along with ReLU activation function. The final layer maps the resulting latent representation to a two-dimensional output vector corresponding to the target classes, namely BL Lac objects and FSRQs. A schematic illustration of the network architecture is shown in Figure~\ref{fig:fnn-arc}.

Model training was performed by minimizing the cross-entropy loss function \citep{mao2023crossentropylossfunctionstheoretical}, which is commonly used for probabilistic classification problems. Optimization was carried out using the AdamW optimizer \citep{kingma2017adammethodstochasticoptimization}, with a decoupled weight decay of $1\times10^{-5}$ to improve regularization. The learning rate was set to $1\times10^{-4}$. 

Training was conducted using a mini-batch size of 32 strategy for a total of 700 epochs. We note that the network operates on static, catalog-level features rather than raw time-series (light-curve) data. Consequently, the model learns from long-term averaged properties that implicitly capture variability and spectral characteristics encoded in the catalog measurements.

To ensure statistical robustness and mitigate biases arising from arbitrary data partitioning, we implemented a 5-fold cross-validation strategy \citep{gorriz2024kfoldcrossvalidationbest}. For each fold, the combined dataset was split into 80\% for training and 20\% for validation. Performance metrics, including loss and accuracy, were monitored on the validation subset after every epoch to track generalization and detect potential overfitting.

\section{Results and Discussions} \label{result_and _disc}

\subsection{Feature Importance Analysis}

Before evaluating the classification performance of the Artificial Neural Network, it is essential to quantify the discriminative power of the individual features selected for this study. To achieve this, we employed Mutual Information (MI) \citep{CoverThomas}, a non-parametric statistical method that measures the dependence between two random variables. Unlike correlation coefficients, which capture only linear relationships, MI quantifies the amount of information obtained about one random variable (the target class: BL Lac or FSRQ) through observing the other (the input feature). Formally, the mutual information $I(X;Y)$ between a feature $X$ and the class label $Y$ is defined as:

\begin{equation}
    I(X;Y) = \sum_{y \in Y} \sum_{x \in X} p(x,y) \log{ \left( \frac{p(x,y)}{p(x)p(y)} \right) }
\end{equation}

where $p(x,y)$ is the joint probability distribution, and $p(x)$ and $p(y)$ are the marginal distributions. A higher MI value indicates a stronger reduction in uncertainty about the target class given the feature.

\begin{figure}[htbp!]
    \centering
    \includegraphics[width=0.9\linewidth]{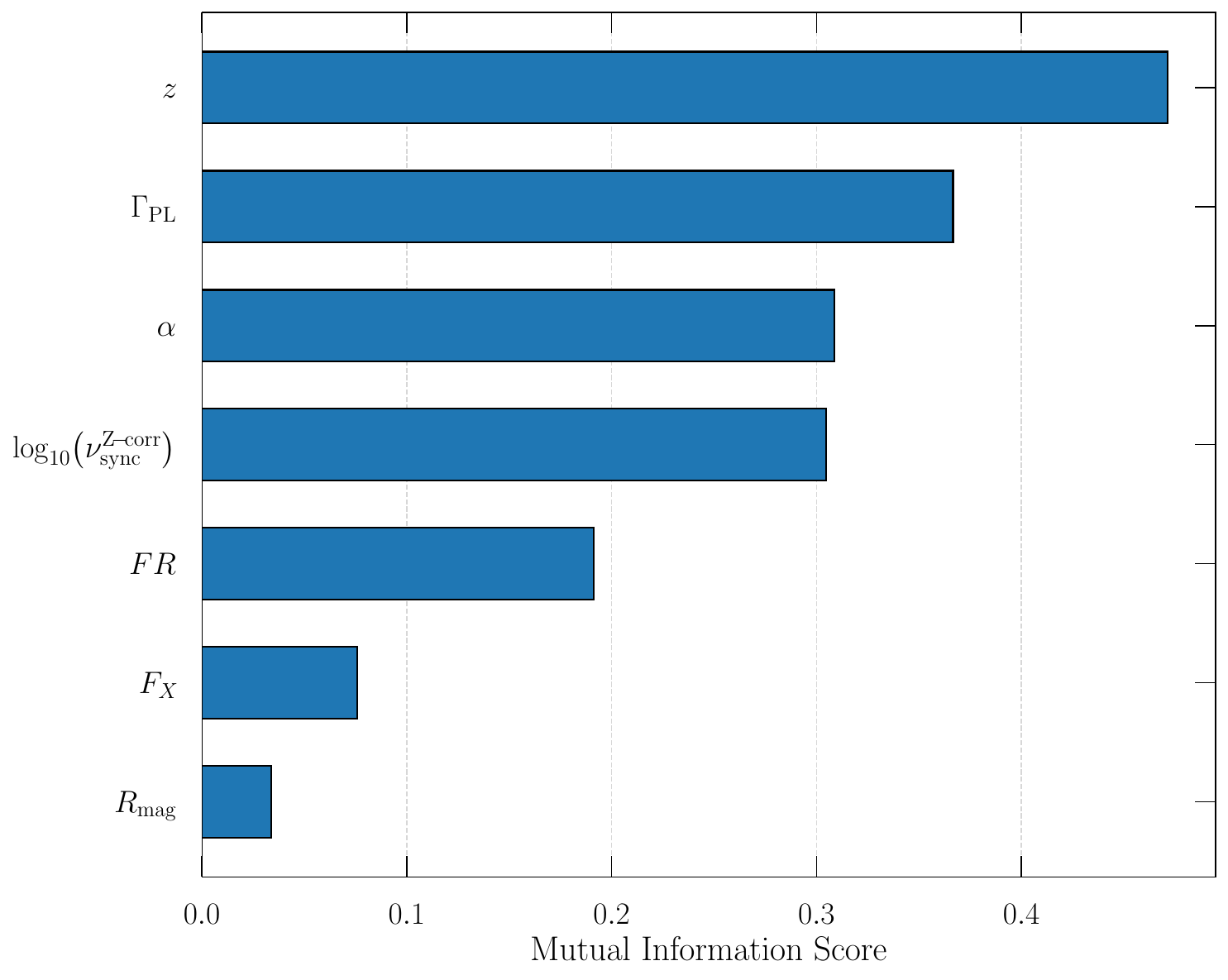}
    \caption{Mutual Information scores quantifying the dependency between input features and blazar subclass labels. The features include redshift ($z$), Power-Law Index ($\Gamma_{\rm PL}$), Spectral Index ($\alpha$), Radio Flux ($F_R$), X-ray Flux ($F_X$), and Optical Magnitude ($R_{\rm mag}$). Higher values indicate greater predictive power regarding the source classification (BL Lac vs. FSRQ).}
    \label{fig:MI}
\end{figure}

Figure \ref{fig:MI} presents the estimated Mutual Information scores for the features utilized in Dataset II (which encompasses the full spectral information). The analysis reveals a distinct hierarchy in feature relevance. The redshift ($z$) exhibits the highest mutual information score, confirming its status as the most critical discriminator between FSRQs and BL Lacs. This is consistent with astrophysical expectations, as FSRQs are generally high-redshift sources with strong emission lines, whereas BL Lacs are typically observed at lower redshifts with featureless continua.

Following redshift, the $\gamma$-ray spectral properties, specifically the Power-Law Index ($\Gamma_{\rm PL}$) and the Spectral Index ($\alpha$), show significant discriminative value. This reflects the known dichotomy in the $\gamma$-ray spectra of blazars, where FSRQs typically display softer spectra (larger photon indices) compared to the harder spectra often seen in BL Lacs. This is in agreement with \citep{2020ApJ...892..105A}.

The multi-wavelength flux measurements, Radio flux ($F_R$), X-ray flux ($F_X$), and Optical magnitude ($R_{\rm mag}$), contribute a moderate but non-negligible amount of information. While their individual scores are lower than that of redshift or spectral indices, their inclusion provides complementary information regarding the shape and normalization of the Spectral Energy Distribution (SED). The combination of these weaker individual predictors within a non-linear framework (such as an ANN) is crucial for resolving edge cases where redshift or spectral indices alone may be ambiguous.

\subsection{Model Evaluation}

We now present the classification performance of the ANN model evaluated using a $k$-fold cross-validation procedure with five folds. In this approach, the full dataset is partitioned into five mutually exclusive subsets of approximately equal size. In each iteration, four subsets are used for training while the remaining subset is held out for testing, and this process is repeated until each subset has served as the test set exactly once. Unless otherwise stated, all reported metrics represent averages computed over the held-out test subsets from each fold.

The use of $k$-fold cross-validation is particularly important in machine-learning applications, as it provides a robust estimate of model performance by reducing sensitivity to any single train–test split. This is especially relevant in our case, where the available labeled sample size is limited and the underlying class distribution between BL Lac objects and FSRQs is moderately imbalanced. By evaluating the model across multiple data partitions, we ensure that the reported performance reflects genuine generalization capability rather than chance alignment with a specific data split.

To further mitigate potential biases arising from class imbalance, we assess model performance using class-sensitive metrics, including precision, recall, and the F1-score, in addition to overall accuracy and confusion matrices. Together, this evaluation strategy provides a comprehensive and reliable assessment of the ANN’s ability to distinguish between the two blazar subclasses.

The training and evaluation sets were constructed by cross-matching the ROMA-BZCAT catalogue with 4FGL-DR4 (Dataset~1) and 3LAC (Dataset~2). For Dataset~1, applying the filter yielded a total of 1,499 sources (809 BL Lacs, 690 FSRQs). This dataset was partitioned into an 80-20\% train–test split, resulting in 1,199 training samples and 300 test samples. Similarly, Dataset~2 and 3, yielded 988 sources (567 BL Lacs, 421 FSRQs), partitioned into 790 training and 198 test samples.

\begin{figure*}[htbp!]
    \centering
    \begin{subfigure}[t]{0.48\linewidth}
        \centering
        \includegraphics[width=0.9\linewidth]{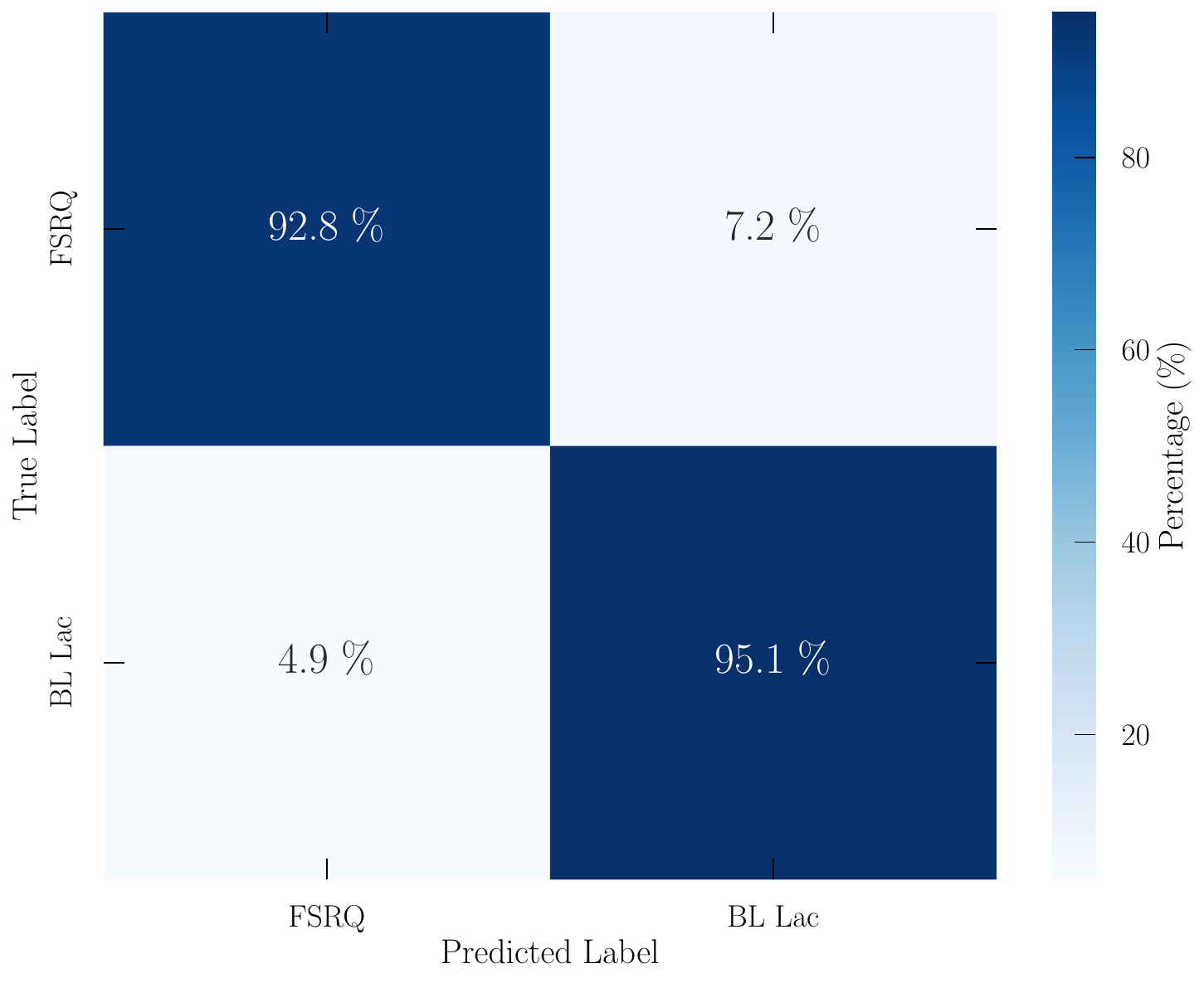}
        \caption{Dataset 1: Confusion matrix based on radio flux, optical magnitude, X-ray flux, and redshift.}
        \label{fig:d1cm}
    \end{subfigure}
    \hfill
    \begin{subfigure}[t]{0.48\linewidth}
        \centering
        \includegraphics[width=0.9\linewidth]{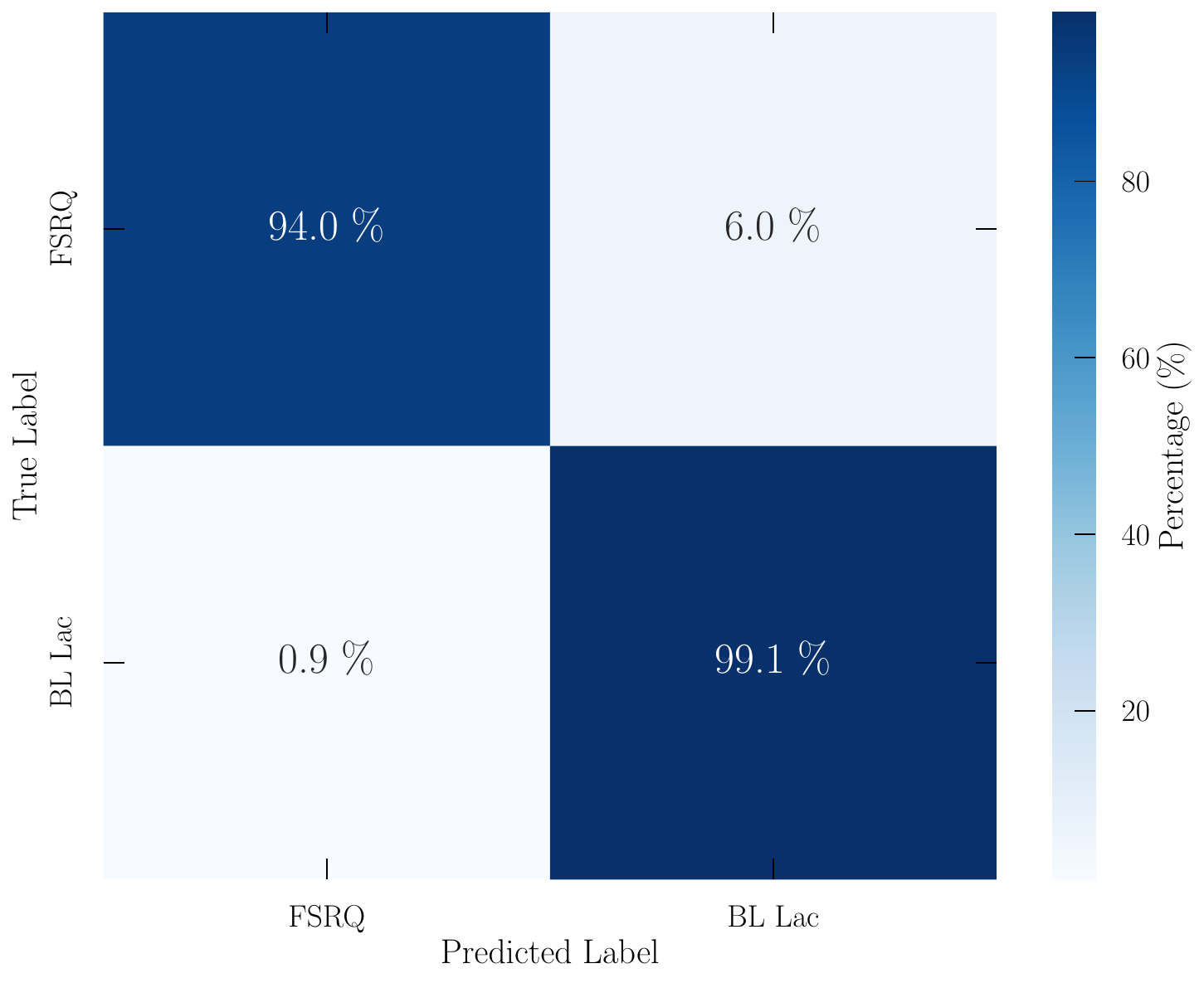}
        \caption{Dataset 2: Confusion matrix with features such as redshift, synchrotron peak frequency, spectral index, and power-law index.}
        \label{fig:d2cm}
    \end{subfigure}
    \hfill
    \begin{subfigure}[t]{0.48\linewidth}
        \centering
        \includegraphics[width=0.9\linewidth]{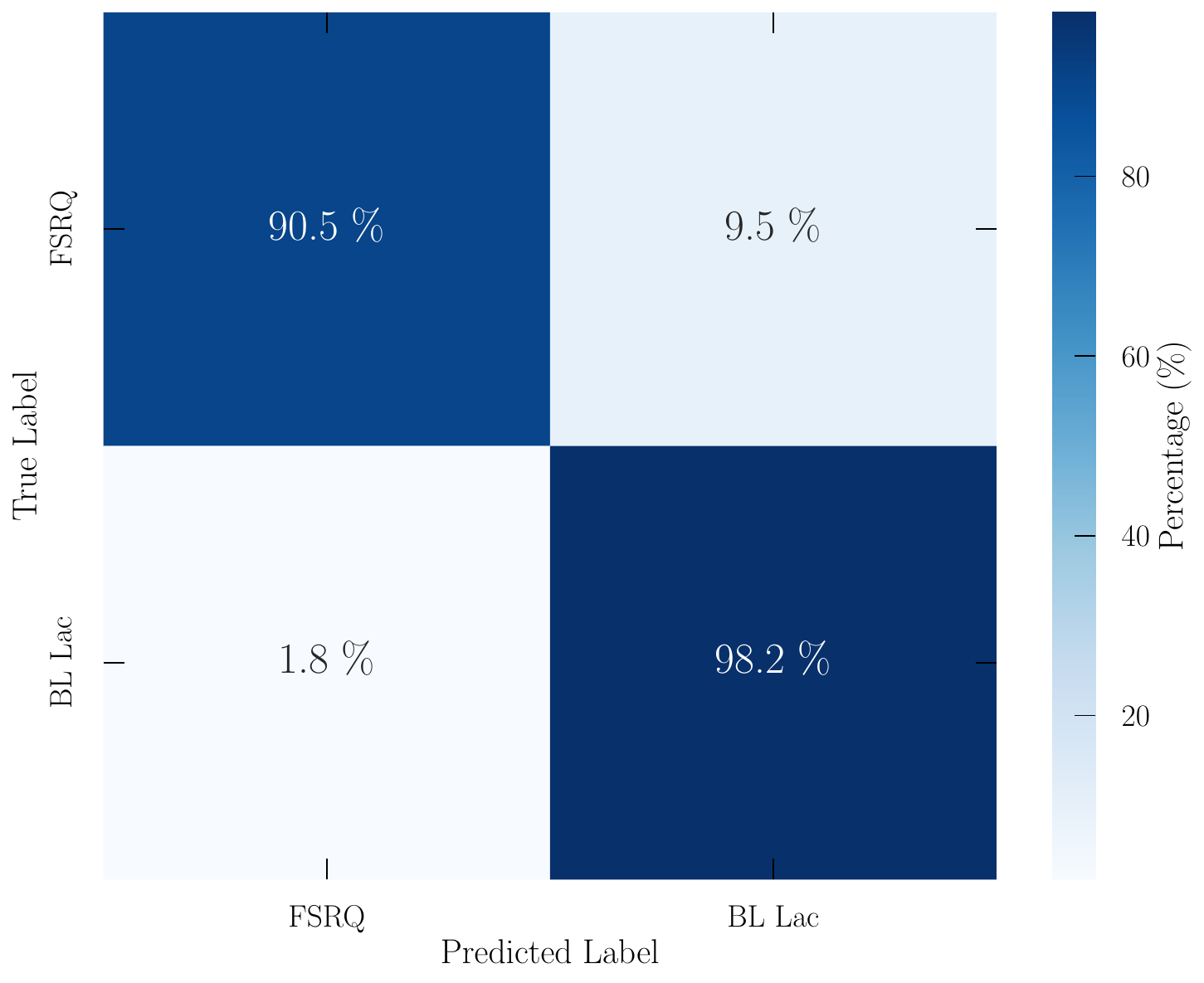}
        \caption{Dataset 3: Confusion matrix with all 7 features.}
        \label{fig:d3cm}
    \end{subfigure}
    \caption{Comparison of confusion matrices showing classification performance for the two datasets.  Subfigure (a) shows the confusion matrix plot for Dataset 1, and Subfigure (b) presents the confusion matrix plot for Dataset 2. All values are normalized by the total number of sources in each true class and expressed as percentages for clarity.
}
    \label{fig:confusion_matrices}
\end{figure*}

Figures~\ref{fig:d1cm}, \ref{fig:d2cm}, and \ref{fig:d3cm} show the confusion matrices obtained for Dataset~1, Dataset~2, and Dataset~3, respectively, by taking the best result out of all 5 folds. For Dataset~1, which is based on radio flux, optical magnitude, X-ray flux, and redshift, the classifier demonstrates stable and well-balanced performance. The confusion matrix indicates that 90.8\% of true FSRQs and 92.4\% of true BL Lac objects are correctly classified. The misclassification rates remain below 10\% for both classes, confirming that even a minimal set of broadband observables contains sufficient information to separate the two blazar populations with reasonable accuracy.

A clear improvement is observed for Dataset~2 (Figure~\ref{fig:d2cm}), which incorporates redshift and physically motivated spectral parameters, including the synchrotron peak frequency, spectral index, and power-law index. In this case, the classifier correctly identifies 94.4\% of FSRQs and 95.4\% of BL Lac objects, yielding the most balanced and overall best performance among the three datasets. The reduced off-diagonal elements indicate that confusion between the two subclasses is minimized when these features are used together. This result highlights that spectral properties, which are directly linked to the underlying emission mechanisms and particle energy distributions, provide a cleaner separation between BL Lacs and FSRQs than flux-based measurements alone. The strong performance of Dataset~2 therefore reflects the fact that these parameters capture intrinsic physical differences between the two classes, rather than observational or distance-related effects.

The confusion matrix for Dataset~3 (Figure~\ref{fig:d3cm}), which combines all seven features, shows a more asymmetric behavior. While the classification of BL Lac objects remains high at 95.4\%, the correct identification rate for FSRQs drops to 85.4\%, with a corresponding increase in confusion toward the BL Lac class. This suggests that adding multi-wavelength flux measurements to an already informative spectral feature set does not necessarily enhance performance. Instead, the inclusion of lower-information features appears to dilute the discriminative power of the spectral parameters, leading to increased overlap in the feature space.

These findings are strongly supported by the mutual information analysis shown in Figure~\ref{fig:MI}. Redshift emerges as the most informative feature, followed by the power-law index, spectral index, and synchrotron peak frequency, all of which play a central role in Dataset~2. In contrast, radio flux, X-ray flux, and optical magnitude exhibit significantly lower mutual information scores, indicating limited standalone discriminative power. The superior performance of Dataset~2 can therefore be directly attributed to its focus on high-information, physically meaningful parameters that best encode the intrinsic differences between blazar subclasses. Overall, this demonstrates that careful feature selection grounded in physical insight is more effective than simply increasing the dimensionality of the input space.

To provide a more detailed assessment of the model performance, we computed precision, recall, and F1-scores for one of the validation fold. The corresponding classification reports for Dataset~1, Dataset~2, and Dataset~3 are presented in Tables~\ref{tab:dataset1_metrics}, \ref{tab:dataset2_metrics}, and \ref{tab:dataset3_metrics}, respectively.

For Dataset~1 (Table~\ref{tab:dataset1_metrics}), the model achieves an overall accuracy of 0.92 on the evaluated sample. Both FSRQs and BL Lac objects are classified with high and well-balanced performance, as reflected by comparable precision, recall, and F1-scores for the two classes. The F1-scores of 0.91 for FSRQs and 0.92 for BL Lac objects indicate that the model does not exhibit a strong bias toward either class when trained on the four-feature multi-wavelength set.

\begin{table}[h!]
\centering
\caption{Classification report for Dataset~1 (FSRQ vs BL Lac).}
\label{tab:dataset1_metrics}
\begin{tabular}{lcccc}
\hline
\textbf{Class} & \textbf{Precision} & \textbf{Recall} & \textbf{F1-score} & \textbf{Support} \\
\hline
FSRQ   & 0.91 & 0.91 & 0.91 & 142 \\
BL Lac & 0.92 & 0.92 & 0.92 & 158 \\
\hline
\textbf{Accuracy} & \multicolumn{4}{c}{0.92 (on 300 samples)} \\
\textbf{Macro Avg} & 0.92 & 0.92 & 0.92 & 300 \\
\textbf{Weighted Avg} & 0.92 & 0.92 & 0.92 & 300 \\
\hline
\end{tabular}
\end{table}

The classification results for Dataset~2 are summarized in Table~\ref{tab:dataset2_metrics}. An overall accuracy of 0.95 is achieved, with strong and balanced performance across both classes. The recall for the FSRQ class is 0.94, indicating that most FSRQs are correctly identified when using spectral features such as the synchrotron peak frequency and spectral indices. The BL Lac class also shows robust performance, with an F1-score of 0.95. The close agreement between precision and recall for both classes suggests that the spectral features provide strong and symmetric discriminative power for distinguishing between FSRQs and BL Lac objects.

\begin{table}[h!]
\centering
\caption{Classification report for Dataset~2 (FSRQ vs BL Lac).}
\label{tab:dataset2_metrics}
\begin{tabular}{lcccc}
\hline
\textbf{Class} & \textbf{Precision} & \textbf{Recall} & \textbf{F1-score} & \textbf{Support} \\
\hline
FSRQ   & 0.94 & 0.94 & 0.94 & 87 \\
BL Lac & 0.95 & 0.95 & 0.95 & 111 \\
\hline
\textbf{Accuracy} & \multicolumn{4}{c}{0.95 (on 198 samples)} \\
\textbf{Macro Avg} & 0.95 & 0.95 & 0.95 & 198 \\
\textbf{Weighted Avg} & 0.95 & 0.95 & 0.95 & 198 \\
\hline
\end{tabular}
\end{table}

Table~\ref{tab:dataset3_metrics} presents the classification report for Dataset~3, which combines both multi-wavelength flux measurements and spectral parameters. The model attains an overall accuracy of 0.91, which is slightly lower than that obtained for Dataset~2. The recall for FSRQs is 0.85, while the BL Lac class maintains a strong F1-score of 0.92. These results indicate that, although the combined feature set preserves reasonable classification performance, the inclusion of both spectral and broadband flux information does not lead to a clear improvement and may introduce additional complexity that modestly affects FSRQ recovery.

\begin{table}[h!]
\centering
\caption{Classification report for Dataset~3 (FSRQ vs BL Lac).}
\label{tab:dataset3_metrics}
\begin{tabular}{lcccc}
\hline
\textbf{Class} & \textbf{Precision} & \textbf{Recall} & \textbf{F1-score} & \textbf{Support} \\
\hline
FSRQ   & 0.94 & 0.85 & 0.89 & 87 \\
BL Lac & 0.89 & 0.95 & 0.92 & 111 \\
\hline
\textbf{Accuracy} & \multicolumn{4}{c}{0.91 (on 198 samples)} \\
\textbf{Macro Avg} & 0.91 & 0.90 & 0.91 & 198 \\
\textbf{Weighted Avg} & 0.91 & 0.91 & 0.91 & 198 \\
\hline
\end{tabular}
\end{table}

Overall, the metrics reported across the three datasets reinforce the conclusions drawn from the confusion matrices, with Dataset~2 emerging as the most effective feature combination. While all datasets yield strong classification performance, Dataset~2 consistently achieves the highest accuracy and the most balanced precision–recall behavior across both classes. This indicates that the spectral feature set used in Dataset~2 captures the intrinsic physical differences between FSRQs and BL Lac objects more effectively than broadband flux-based features alone or their direct combination. In contrast, Datasets~1 and~3 show slightly increased class asymmetries, reflecting how different feature choices emphasize distinct physical characteristics of blazars and can influence classification balance.

\begin{figure*}[htbp!]
    \centering
    \begin{subfigure}[t]{0.48\linewidth}
        \centering
        \includegraphics[width=0.9\linewidth]{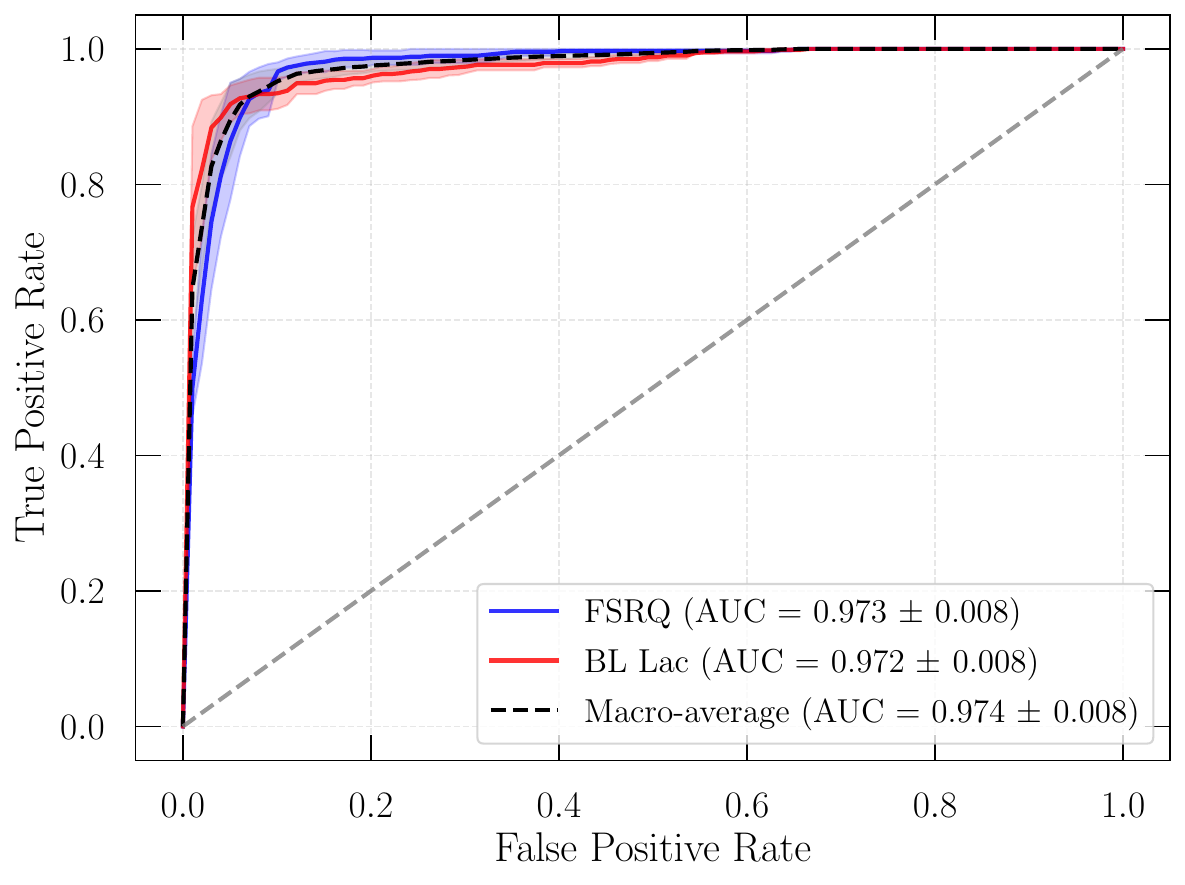}
        \caption{ROC curve for Dataset 1.}
        \label{fig:roc1}
    \end{subfigure}
    \hfill
    \begin{subfigure}[t]{0.48\linewidth}
        \centering
        \includegraphics[width=0.9\linewidth]{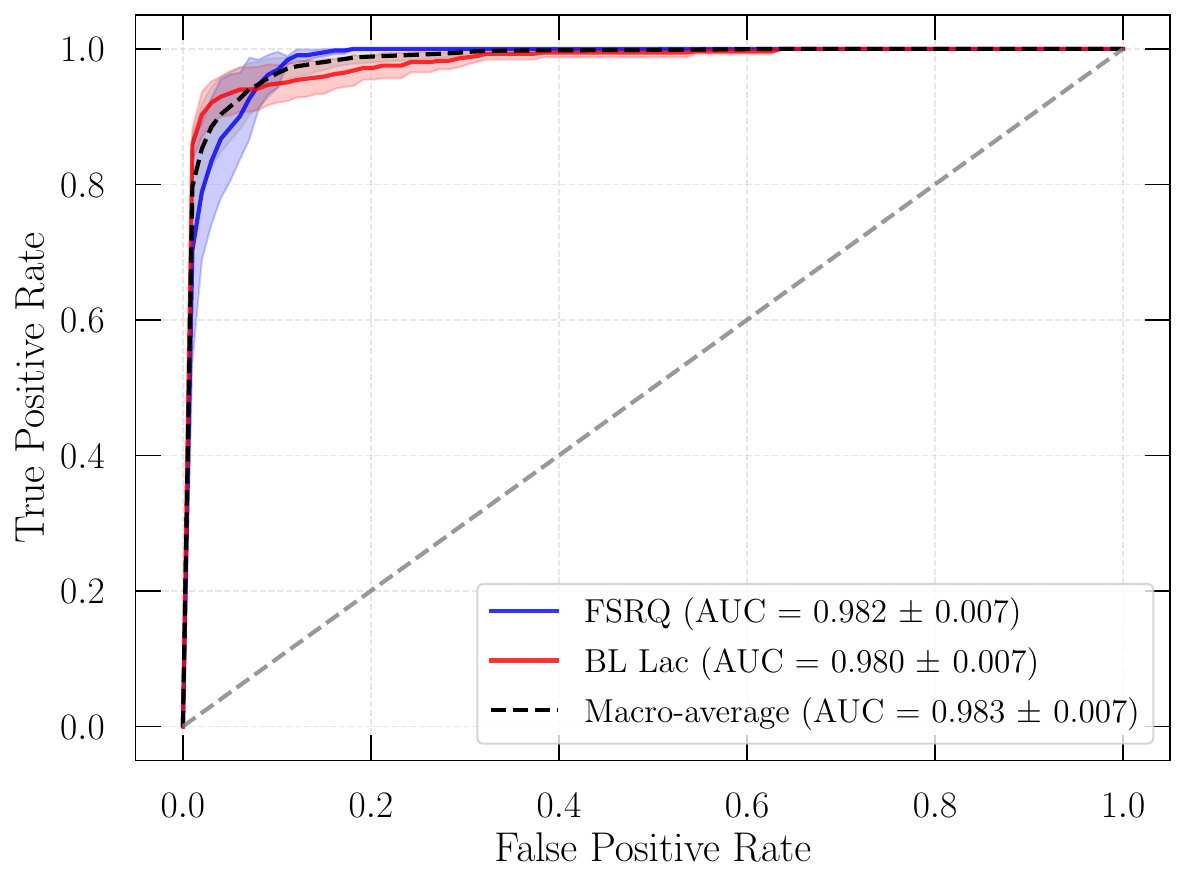}
        \caption{ROC curve for Dataset 2.}
        \label{fig:roc2}
    \end{subfigure}
    \hfill
    \begin{subfigure}[t]{0.48\linewidth}
        \centering
        \includegraphics[width=0.9\linewidth]{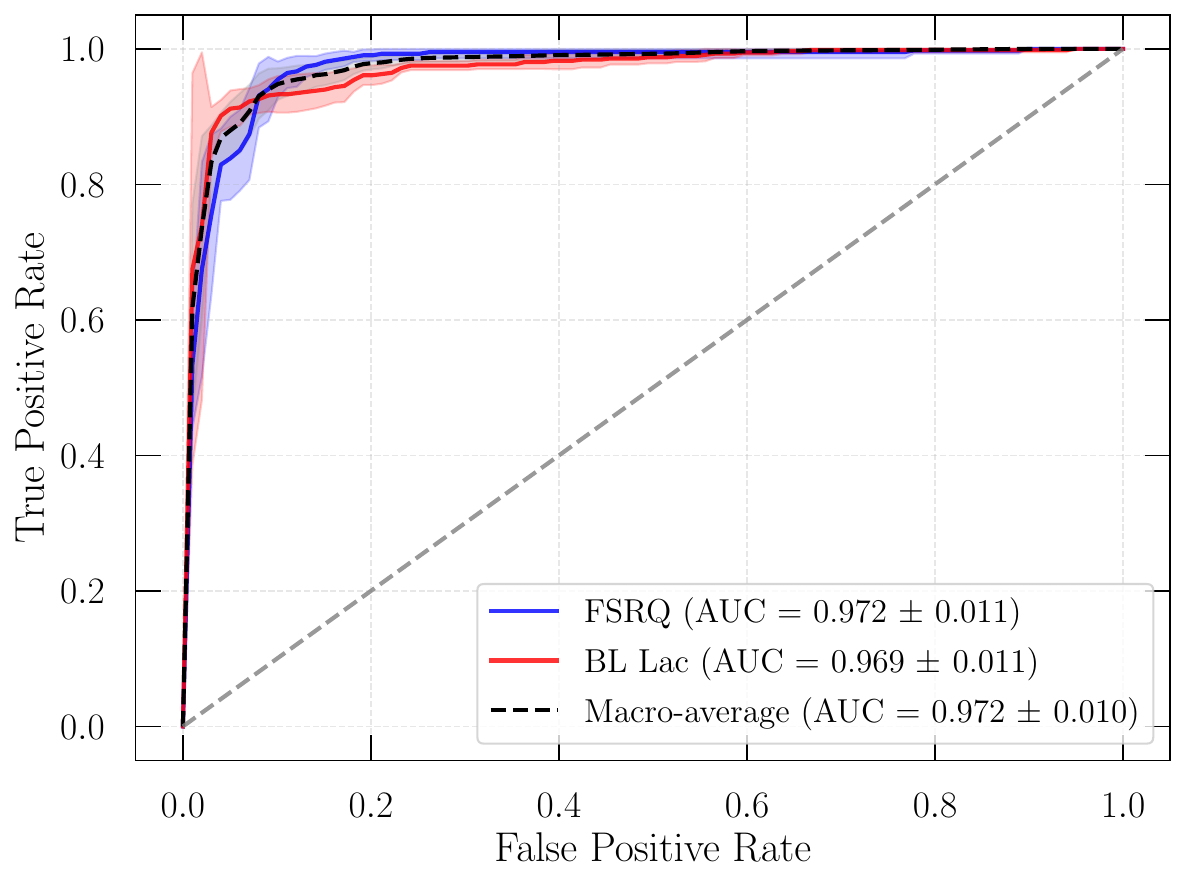}
        \caption{ROC curve for Dataset 3.}
        \label{fig:roc3}
    \end{subfigure}
    \caption{Receiver Operating Characteristic (ROC) curves illustrating the performance of the classification model on two independent datasets. Subfigure (a) shows the ROC curve for Dataset 1, and Subfigure (b) presents the ROC curve for Dataset 2, allowing direct comparison of model effectiveness across different datasets.}
    \label{fig:roc_curves}
\end{figure*}

We further validated the model’s discriminative capability using Receiver Operating Characteristic (ROC) and Precision–Recall (PR) curves. Figures~\ref{fig:roc1}, \ref{fig:roc2}, and \ref{fig:roc3} show the ROC curves for Dataset~1, Dataset~2, and Dataset~3, respectively. In each case, the displayed curve corresponds to the mean ROC computed across the $k$ cross-validation folds, while the shaded region represents one standard deviation. The strong overlap of the ROC curves across folds and the narrow uncertainty bands indicate that the classification performance is stable and largely insensitive to the specific data partitioning. The consistently high Area Under the Curve (AUC) values across all datasets confirm strong class separability. These findings are consistent with previous deep-learning studies in related contexts, such as \cite{2021MNRAS.505.1268F}, which demonstrated similarly stable AUC behavior across multiple cross-validation runs when using broadband spectral energy distributions for AGN classification.

\begin{figure*}[htbp!]
    \centering
    \begin{subfigure}[t]{0.48\linewidth}
        \centering
        \includegraphics[width=0.9\linewidth]{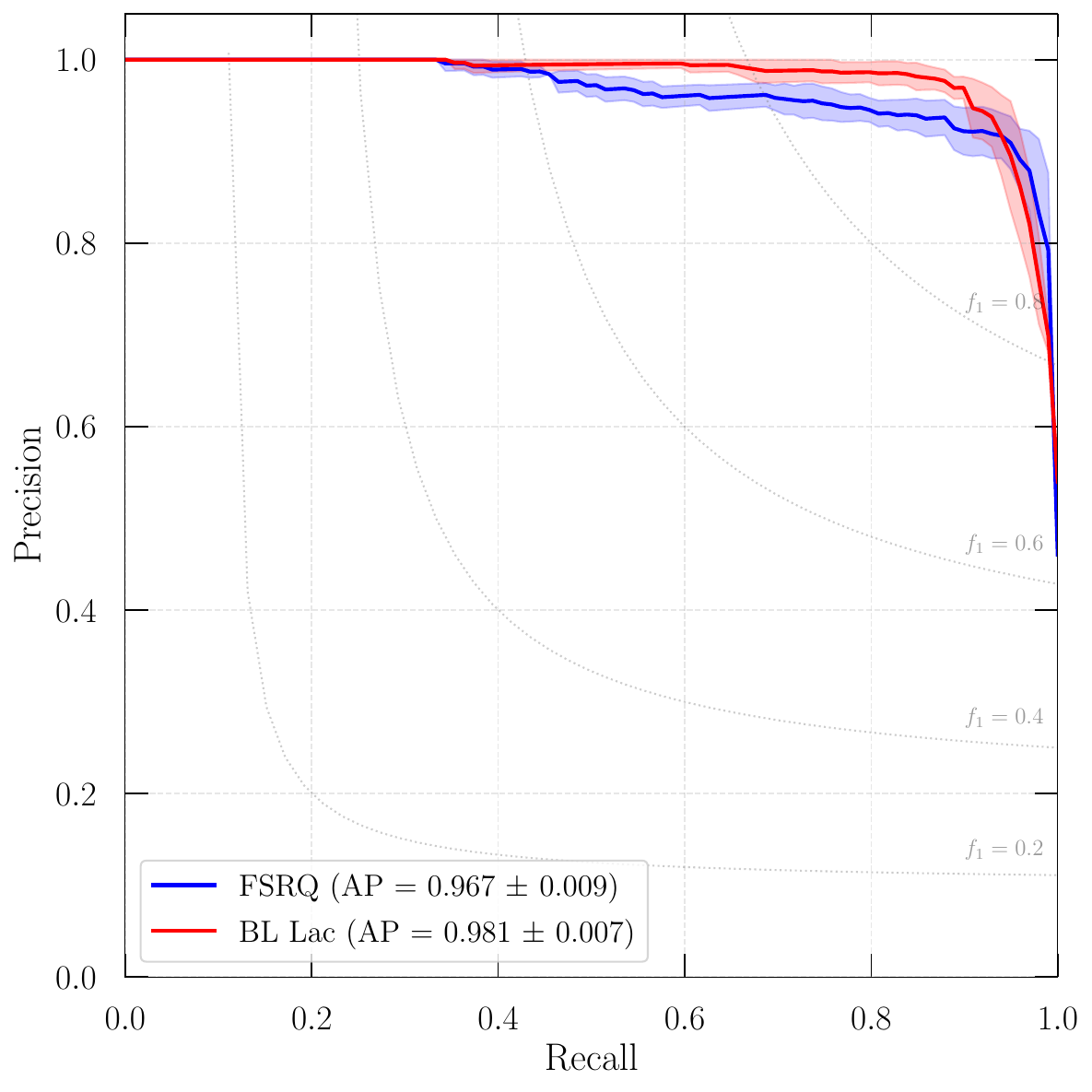}
        \caption{Precision-recall curve for Dataset 1.}
        \label{fig:pr1}
    \end{subfigure}
    \hfill
    \begin{subfigure}[t]{0.48\linewidth}
        \centering
        \includegraphics[width=0.9\linewidth]{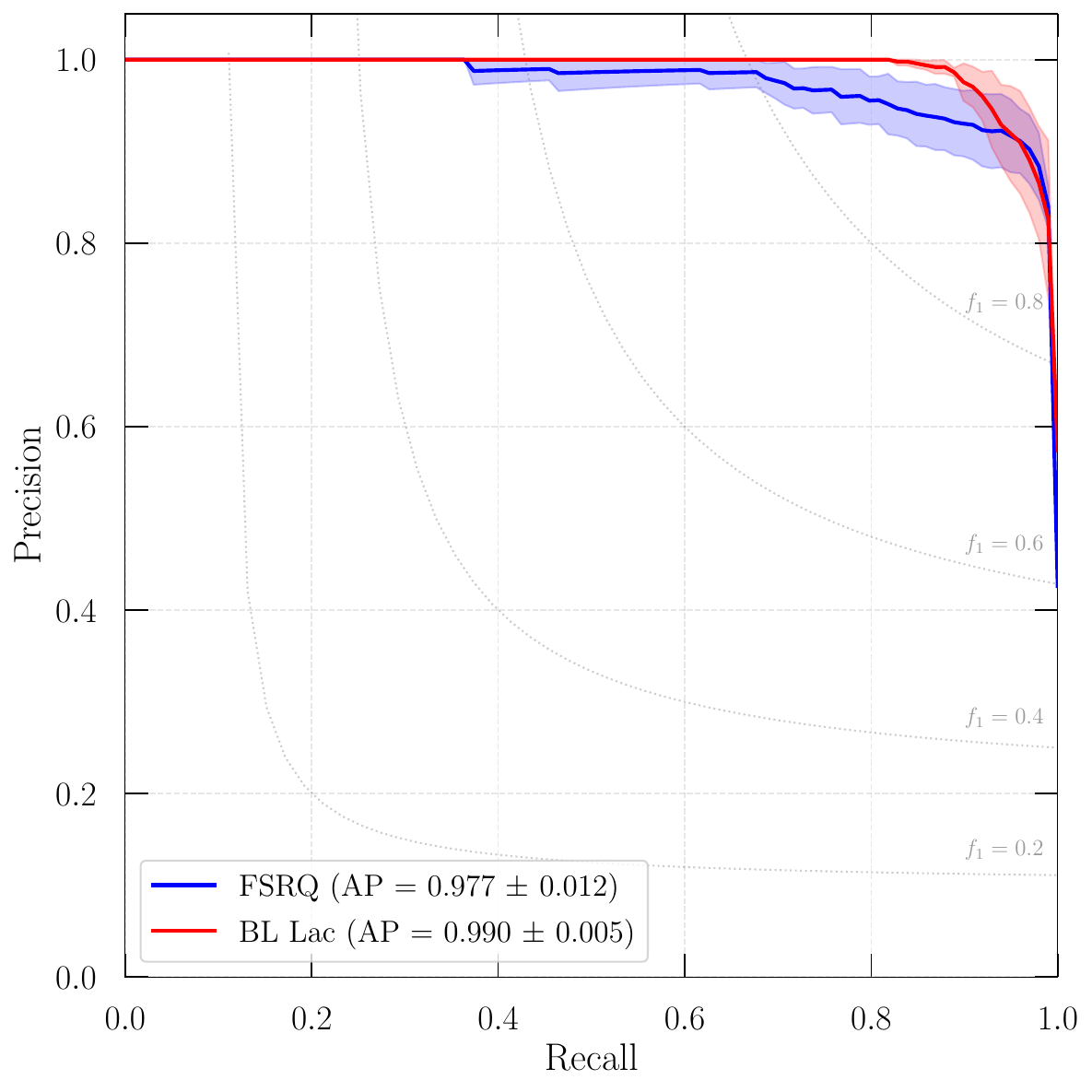}
        \caption{Precision-recall curve for Dataset 2.}
        \label{fig:pr2}
    \end{subfigure}
    \hfill
    \begin{subfigure}[t]{0.48\linewidth}
        \centering
        \includegraphics[width=0.9\linewidth]{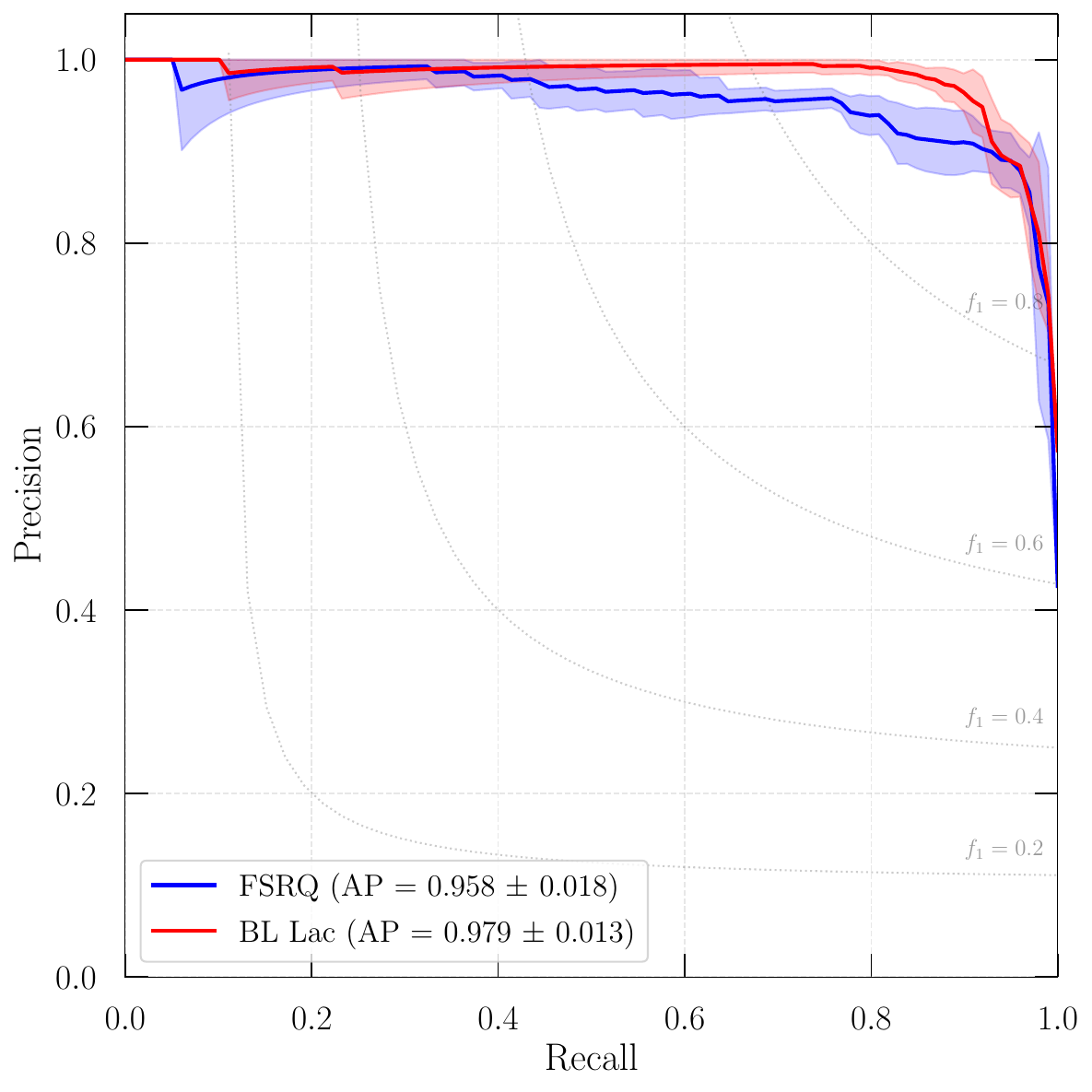}
        \caption{Precision-recall curve for Dataset 3.}
        \label{fig:pr3}
    \end{subfigure}
    \caption{Precision-recall (PR) curves illustrating the performance of the classification model on two independent datasets. Subfigure (a) shows the PR curve for Dataset 1, and Subfigure (b) presents the (PR) curve for Dataset 2, allowing direct comparison of model effectiveness across different datasets.}

    \label{fig:pr_curves}
\end{figure*}

Given the moderate class imbalance, the precision–recall (PR) curves shown in Figures~\ref{fig:pr1}, \ref{fig:pr2}, and \ref{fig:pr3} provide a complementary and more sensitive assessment of performance. As with the ROC analysis, the PR curves represent the mean behavior across the $k$ folds, with shaded regions indicating the corresponding standard deviation. The close agreement between folds demonstrates that the precision–recall trade-off is robust against variations in the training–validation split. Across all three datasets, the PR curves remain near the upper-right region of the plane, indicating high precision over a broad range of recall values, with a noticeable decline only as recall approaches unity. Among the three cases, Dataset~2 consistently exhibits the most favorable PR characteristics, achieving higher precision at comparable recall levels and showing the smallest inter-fold variability, further highlighting its balanced and reliable classification performance.

\begin{figure*}[htbp!]
    \centering
    \begin{subfigure}[t]{0.48\linewidth}
        \centering
        \includegraphics[width=0.9\linewidth]{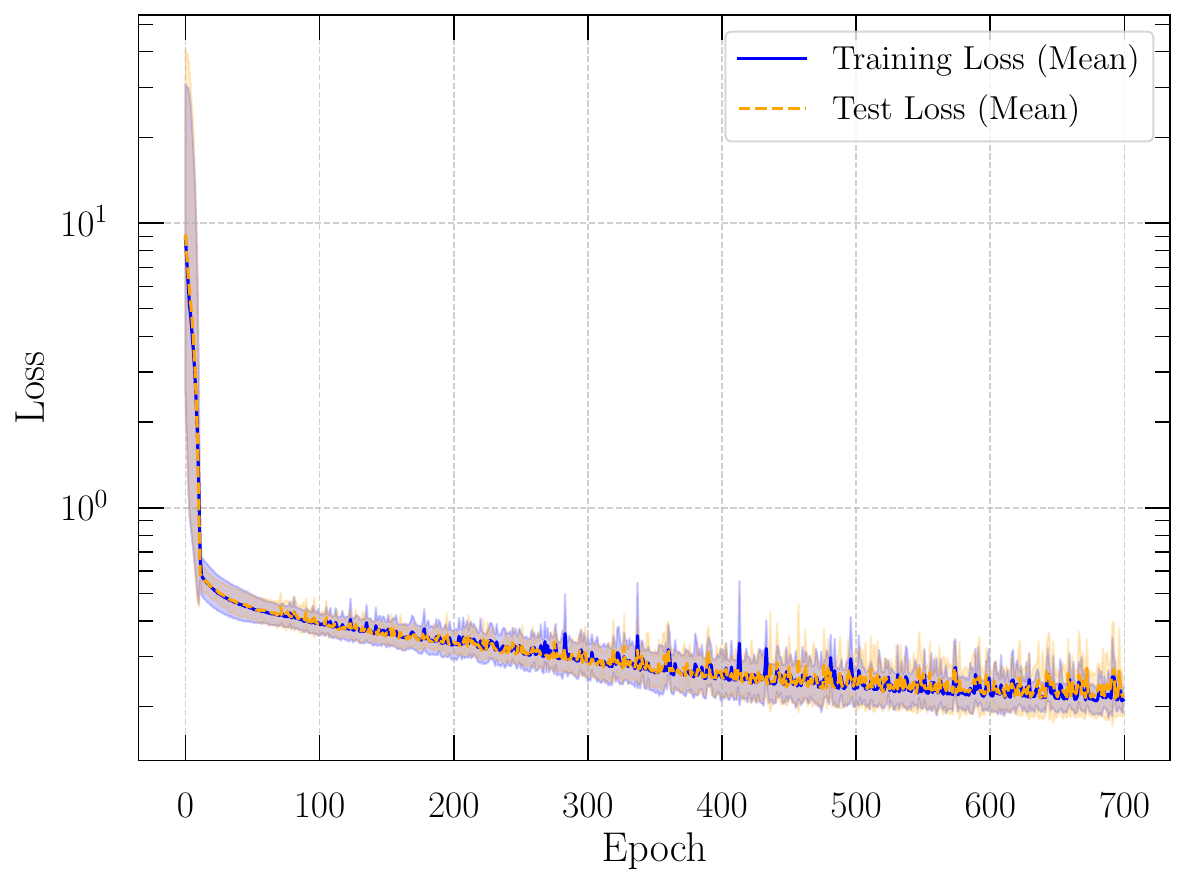}
        \caption{Training and validation loss curves for Dataset 1.}
        \label{fig:loss1}
    \end{subfigure}
    \hfill
    \begin{subfigure}[t]{0.48\linewidth}
        \centering
        \includegraphics[width=0.9\linewidth]{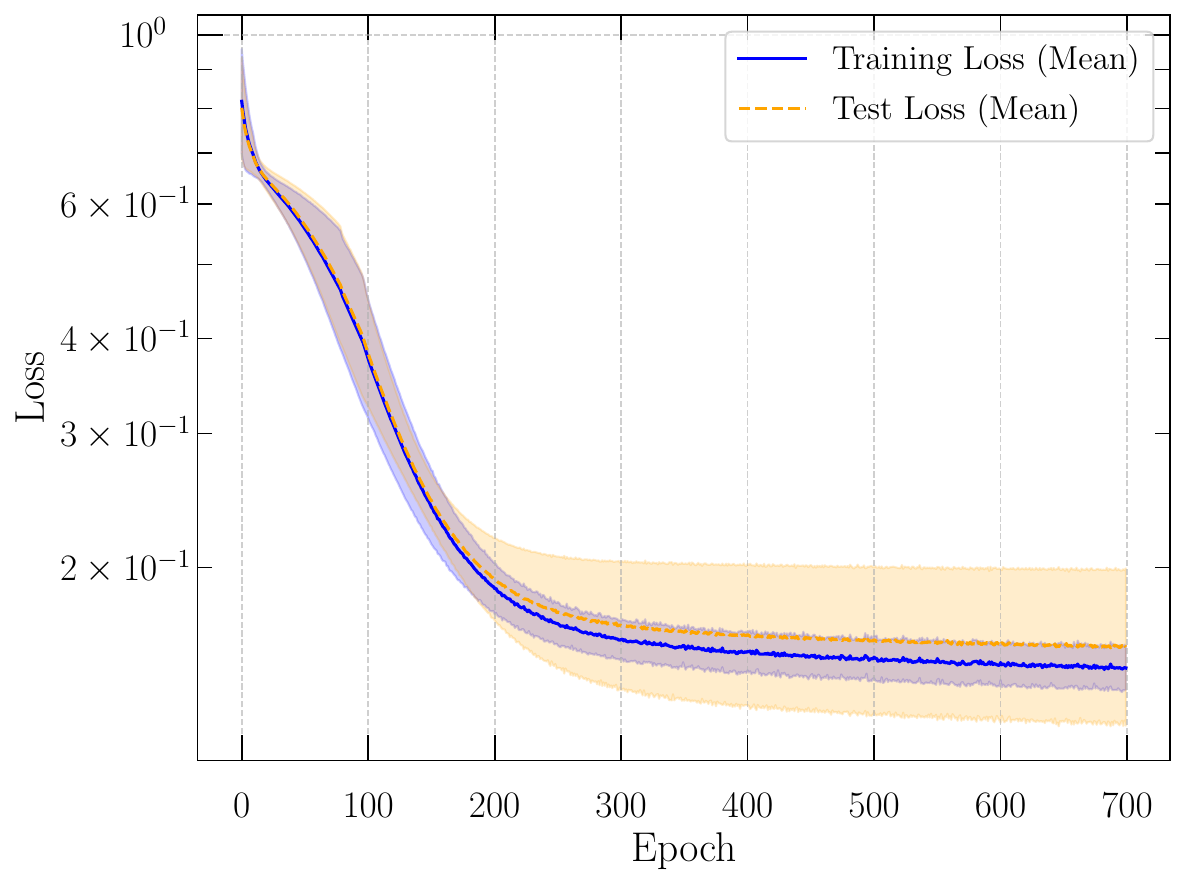}
        \caption{Training and validation loss curves for Dataset 2.}
        \label{fig:loss2}
    \end{subfigure}
    \hfill
    \begin{subfigure}[t]{0.48\linewidth}
        \centering
        \includegraphics[width=0.9\linewidth]{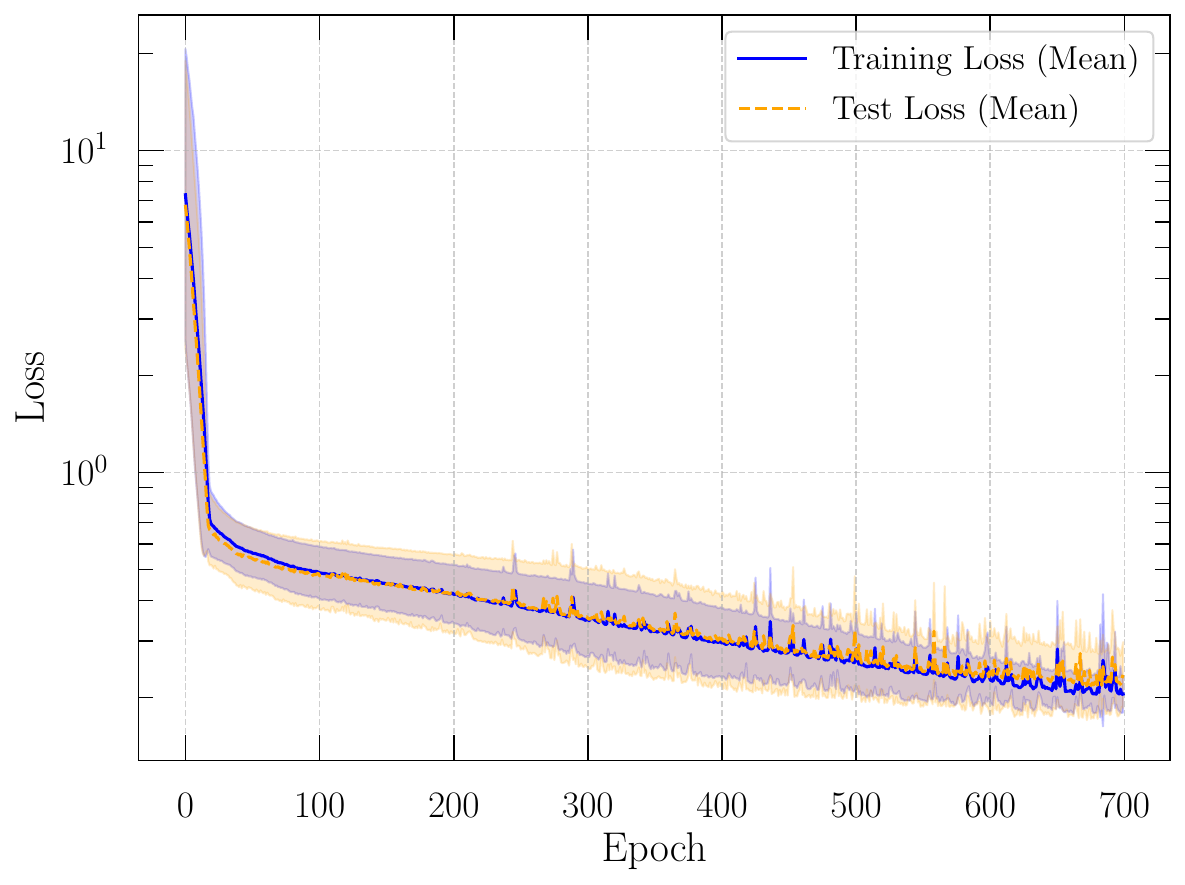}
        \caption{Training and validation loss curves for Dataset 3.}
        \label{fig:loss3}
    \end{subfigure}
    \caption{Loss curve comparison indicating the convergence behavior for the two datasets.}
    \label{fig:loss_curves}
\end{figure*}

The training and validation loss curves shown in Figures~\ref{fig:loss1}, \ref{fig:loss2}, and \ref{fig:loss3} provide additional evidence for the robustness of the learning process under $k$-fold cross-validation. For all datasets, both training and validation losses decrease smoothly over epochs, with no persistent divergence between the two, indicating the absence of overfitting or underfitting. When examined across folds, Dataset~2 again stands out by exhibiting consistently smooth and stable convergence behavior, with minimal fluctuations in the validation loss. In contrast, Dataset~1 and Dataset~3 show occasional oscillations across different folds, suggesting a less uniform optimization landscape. This enhanced stability for Dataset~2 reinforces the conclusion that its spectral feature set provides a more coherent and well-conditioned representation, leading to improved generalization across different data splits.

\subsection{BCU Classification and Physical Interpretation}

Following model validation, we applied the \emph{best-performing configuration}, corresponding to Dataset~2, to classify Blazar Candidates of Uncertain type (BCUs). This choice is motivated by the consistently superior and more balanced performance of Dataset~2 across all evaluation metrics, including accuracy, precision-recall behavior, and convergence stability. Using this optimized model, we identified a total of 82 BCU sources for which reliable predictions could be made. The full set of probabilistic predictions is reported in \footnote{The complete table of predicted classes for BCUs is given at the following link: \url{https://github.com/AfrozSaqlain/Light_Curve_of_Blazars_ML/tree/main/data}.}, while a representative subset is shown in Table~\ref{tab:bcu_predictions_compact}.

Direct numerical comparison with previously published BCU classification studies is not straightforward, owing to differences in catalog versions, feature definitions, and selection criteria. Nevertheless, the present results clearly demonstrate that catalog-level spectral features, specifically those characterizing the synchrotron peak and high-energy spectral shape, are sufficient to distinguish between BL Lac objects and FSRQs with high confidence. Importantly, these features implicitly encode long-term variability and radiative properties without requiring access to raw light-curve data. In this sense, our approach complements studies that focus on short-timescale variability, offering an alternative and computationally efficient pathway for BCU classification.

From a physical perspective, the success of the neural network indicates that the model captures non-linear correlations associated with the distinct emission environments of the two blazar subclasses. FSRQs are typically associated with radiatively efficient accretion disks and strong external photon fields, leading to characteristic high-energy spectral signatures. In contrast, BL Lac objects are generally dominated by Synchrotron Self-Compton (SSC) emission and exhibit weaker external radiation fields. The ability of the model to separate BCUs into these two categories supports the interpretation that most BCUs do not constitute a separate physical class, but rather represent standard BL Lac or FSRQ systems for which spectroscopic or multi-wavelength information is incomplete or ambiguous.

An additional advantage of the present framework lies in its probabilistic output. We find that high-confidence predictions, with probabilities close to 0 or 1, are typically associated with sources exhibiting canonical spectral properties consistent with their predicted class. Conversely, lower-confidence predictions tend to correspond to sources with intermediate or atypical feature values, which may reflect transitional systems, low signal-to-noise measurements, or genuinely complex physical states. These cases are particularly interesting from an astrophysical standpoint, as they may probe the boundary between traditional blazar subclasses.

\begin{figure}
    \centering
    \includegraphics[width=0.9\linewidth]{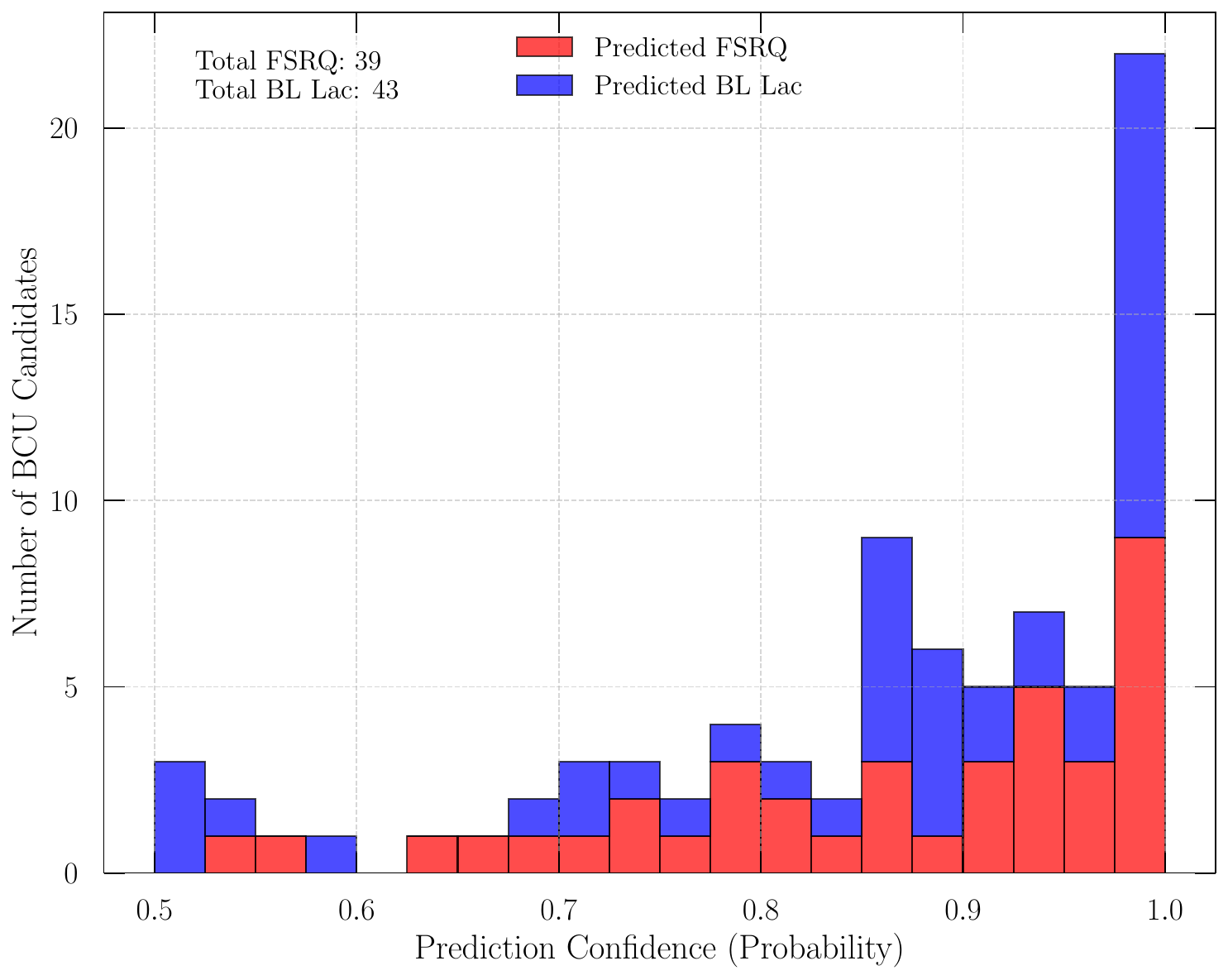}
    \caption{Histogram of confidence in predictions made on BCUs}
    \label{fig:hist}
\end{figure}

Figure~\ref{fig:hist} shows the distribution of prediction confidence for BCUs classified using the best-performing model. The histogram reveals that a large fraction of sources are assigned to either the BL Lac or FSRQ class with high confidence, with probabilities clustered toward values close to unity. This indicates that, for most BCUs, the model identifies clear spectral signatures consistent with one of the two established blazar subclasses. The comparable number of sources predicted as FSRQs and BL Lac objects further suggests that the classification does not suffer from a strong class bias.

At the same time, a smaller subset of BCUs exhibits intermediate confidence values, reflecting increased uncertainty in their classification. These sources likely correspond to objects with less well-defined spectral properties, lower signal-to-noise measurements, or genuinely intermediate physical characteristics. Such cases are of particular astrophysical interest, as they may represent transitional systems or sources requiring improved spectroscopic or multi-wavelength coverage. Overall, the confidence distribution highlights the usefulness of the probabilistic output not only for classification, but also for prioritizing BCUs for targeted follow-up observations.

Overall, the probabilistic classification of BCUs using the best-performing spectral feature set provides not only a robust means of source labeling, but also a physically meaningful ranking of classification confidence. This makes the approach well suited for guiding targeted multi-wavelength follow-up observations aimed at resolving the nature of ambiguous or transitional blazar candidates.

\begin{table*}[htbp!]
    \centering
    \begin{tabularx}{\textwidth}{p{3.8cm} *{7}{X}}
        \hline
        \textbf{Source Name} &
        \textbf{z} &
        \textbf{Log $\nu_{\text{Sync}}$ Z corr. (Hz)} &
        \textbf{Power Law Index} &
        \textbf{Spectral Index} &
        \textbf{Class} &
        \textbf{Predicted Class} &
        \textbf{Prediction Confidence} \\
        \hline

        5BZU J0023+4456 & 2.023 & 13.050 & 2.57 & 2.5685 & BCU & FSRQ & 0.9971 \\
        5BZU J0040+4050 & 0.000 & 0.000 & 1.13 & 1.1323 & BCU & BL Lac & 0.8824 \\
        5BZU J0049-4457 & 0.121 & 13.870 & 2.53 & 2.5279 & BCU & BL Lac & 0.7334 \\
        5BZU J0058+3311 & 1.369 & 13.310 & 2.41 & 2.4073 & BCU & FSRQ & 0.9904 \\
        5BZU J0058-5659 & 0.000 & 12.778 & 2.62 & 2.6164 & BCU & FSRQ & 0.5475 \\
        5BZU J0100+0745 & 0.000 & 13.358 & 1.80 & 1.7966 & BCU & BL Lac & 0.9962 \\
        5BZU J0102+5824 & 0.644 & 12.941 & 2.25 & 2.0943 & BCU & FSRQ & 0.6828 \\
        5BZU J0110+6805 & 0.000 & 14.864 & 1.99 & 1.9909 & BCU & BL Lac & 0.9979 \\
        5BZU J0128+4439 & 0.228 & 13.969 & 2.33 & 2.3320 & BCU & BL Lac & 0.8574 \\
        5BZU J0133-5200 & 0.000 & 12.063 & 2.63 & 2.6276 & BCU & FSRQ & 0.7209 \\

        \hline
    \end{tabularx}

    \caption{ANN-based classification results for Blazar Candidates of Uncertain type (BCUs), listing redshift, synchrotron peak frequency, spectral indices, predicted class labels, and associated confidence scores.}
    \label{tab:bcu_predictions_compact}
\end{table*}

\section{Conclusion} \label{conclusion}
In this work, we developed and evaluated artificial neural network (ANN) models for the classification of blazars using multi-wavelength information drawn from established astronomical catalogues, including the \textit{Fermi}-LAT 4FGL-DR4, ROMA-BZCAT v5, and the 3LAC catalog. Blazars exhibit distinct observational signatures across the electromagnetic spectrum, and our analysis pipeline was designed to exploit this property by incorporating physically motivated features spanning radio, optical, X-ray, and high-energy spectral domains.

Our classification framework, implemented in PyTorch, demonstrates that even a relatively simple feed-forward ANN can robustly distinguish between BL Lac objects and FSRQs when trained on carefully selected catalog-level features. Among the different feature combinations explored, Dataset~2, which emphasizes redshift and high-energy spectral properties such as the synchrotron peak frequency and spectral indices, consistently yielded the best performance. This configuration achieved the most balanced precision–recall behavior, the highest overall classification accuracy, and the most stable convergence during training. These results highlight the importance of selecting physically meaningful features that directly reflect the underlying emission processes in blazar jets, rather than relying solely on broadband flux measurements.

The training and validation loss curves across all datasets show convergence, with no indication of overfitting or underfitting. Notably, the loss evolution for Dataset~2 is the most stable, reinforcing the conclusion that spectral features provide a well-conditioned and informative representation for learning. In addition, the consistent performance observed across $k$-fold cross-validation demonstrates that the model’s predictions are robust to different data partitions and are not driven by a particular train–test split. While the current implementation already performs strongly, future improvements could explore more advanced feature engineering or architectural refinements to further enhance convergence and generalization.

A comparison with recent machine-learning–based studies shows that our results are competitive with the state of the art. In particular, our achieved AUC is comparable to, and in some cases exceeds, values reported in the literature. Similarly, we achieve better F1-scores than some of the other recent classification efforts. Differences in catalog versions, feature selection, and sample construction prevent a direct one-to-one comparison, but overall performance trends are in strong agreement.

We also applied the best-performing model to classify Blazar Candidates of Uncertain type (BCUs). The resulting probabilistic predictions indicate that a large fraction of BCUs can be confidently assigned to either the BL Lac or FSRQ class, supporting the interpretation that BCUs do not represent a distinct physical population but rather reflect incomplete observational coverage. The probabilistic nature of the predictions further enables the identification of low-confidence or ambiguous cases, which are prime targets for future spectroscopic and multi-wavelength follow-up observations.

The primary limitation of the present approach lies in the availability of complete multi-wavelength information. Many BCUs still lack reliable optical or X-ray measurements, which restricts their inclusion in the current framework. As future \textit{Fermi}-LAT data releases and upcoming surveys provide more homogeneous and comprehensive coverage, the applicability and impact of this method will increase substantially. In this context, the flexibility of the ANN framework allows new features to be incorporated seamlessly as additional data become available.

Overall, this study demonstrates the effectiveness of machine-learning techniques for blazar classification and provides a scalable, physically motivated framework for future analyses. By combining robust performance with interpretable, catalog-level features, this approach contributes to improved catalog completeness and offers valuable insights into blazar populations and jet physics. Future work will focus on incorporating interpretability tools, such as SHAP values or permutation-based feature importance, to better quantify the role of individual physical parameters in the BL Lac versus FSRQ separation and to further strengthen the connection between machine-learning predictions and underlying astrophysical processes.

\FloatBarrier

\begin{acknowledgments}
We acknowledge the respective groups or collaborations for developing the FERMI-LAT catalogue, ROMA-BZCAT catalogue, and LAT AGN catalogue.
\end{acknowledgments}

\section*{Data Availability}

All multi-wavelength parameters used to train the ANN models in this work are publicly available from the astronomical catalogues referenced in the text, including ROMA-BZCAT, Fermi-LAT 4FGL-DR4, and the 3LAC release. Random seeds were fixed for all training and cross-validation runs to ensure reproducibility of the reported results. The full codebase used for data preprocessing, model training, and evaluation is openly accessible at our GitHub repository:
\url{https://github.com/AfrozSaqlain/Light_Curve_of_Blazars_ML}.


\facilities{All the data analysis and ML training were performed on a  laptop with NVIDIA RTX 4070 and Intel Core i7 processor, with 32 GB of RAM.}

\software{PyTorch \cite{paszke2019pytorchimperativestylehighperformance},
          Scikit-learn \cite{pedregosa2018scikitlearnmachinelearningpython},
          TOPCAT \cite{Taylor_2017}
          }


\FloatBarrier

\bibliography{sample701}{}

@misc{paszke2019pytorchimperativestylehighperformance,
      title={PyTorch: An Imperative Style, High-Performance Deep Learning Library}, 
      author={Adam Paszke and Sam Gross and Francisco Massa and Adam Lerer and James Bradbury and Gregory Chanan and Trevor Killeen and Zeming Lin and Natalia Gimelshein and Luca Antiga and Alban Desmaison and Andreas Köpf and Edward Yang and Zach DeVito and Martin Raison and Alykhan Tejani and Sasank Chilamkurthy and Benoit Steiner and Lu Fang and Junjie Bai and Soumith Chintala},
      year={2019},
      eprint={1912.01703},
      archivePrefix={arXiv},
      primaryClass={cs.LG},
      url={https://arxiv.org/abs/1912.01703}, 
}

@misc{pedregosa2018scikitlearnmachinelearningpython,
      title={Scikit-learn: Machine Learning in Python}, 
      author={Fabian Pedregosa and Gaël Varoquaux and Alexandre Gramfort and Vincent Michel and Bertrand Thirion and Olivier Grisel and Mathieu Blondel and Andreas Müller and Joel Nothman and Gilles Louppe and Peter Prettenhofer and Ron Weiss and Vincent Dubourg and Jake Vanderplas and Alexandre Passos and David Cournapeau and Matthieu Brucher and Matthieu Perrot and Édouard Duchesnay},
      year={2018},
      eprint={1201.0490},
      archivePrefix={arXiv},
      primaryClass={cs.LG},
      url={https://arxiv.org/abs/1201.0490}, 
}

@ARTICLE{2021MNRAS.505.1268F,
       author = {{Fraga}, Bernardo M.~O. and {Barres de Almeida}, Ulisses and {Bom}, Cl{\'e}cio R. and {Brandt}, Carlos H. and {Giommi}, Paolo and {Schubert}, Patrick and {de Albuquerque}, M{\'a}rcio P.},
        title = "{Deep learning Blazar classification based on multifrequency spectral energy distribution data}",
      journal = {\mnras},
         year = 2021,
        month = jul,
       volume = {505},
       number = {1},
        pages = {1268-1279},
          doi = {10.1093/mnras/stab1349},
archivePrefix = {arXiv},
       eprint = {2012.15340},
 primaryClass = {astro-ph.HE},
       adsurl = {https://ui.adsabs.harvard.edu/abs/2021MNRAS.505.1268F}
}

@ARTICLE{1998MNRAS.299..433F,
       author = {{Fossati}, G. and {Maraschi}, L. and {Celotti}, A. and {Comastri}, A. and {Ghisellini}, G.},
        title = "{A unifying view of the spectral energy distributions of blazars}",
      journal = {\mnras},
         year = 1998,
        month = sep,
       volume = {299},
       number = {2},
        pages = {433-448},
          doi = {10.1046/j.1365-8711.1998.01828.x},
archivePrefix = {arXiv},
       eprint = {astro-ph/9804103},
 primaryClass = {astro-ph},
       adsurl = {https://ui.adsabs.harvard.edu/abs/1998MNRAS.299..433F}
}

@article{Kang:2019cck,
    author = "Kang, Shi-Ju and Li, Enze and Ou, Wujing and Zhu, Kerui and Fan, Jun-Hui and Wu, Qingwen and Yin, Yue",
    title = "{Evaluating the classification of Fermi BCUs from the 4FGL Catalog Using Machine Learning}",
    eprint = "1911.02570",
    archivePrefix = "arXiv",
    primaryClass = "astro-ph.HE",
    doi = "10.3847/1538-4357/ab558b",
    month = "11",
    year = "2019"
}

@article{Kovacevic:2020sly,
    author = "Kova{\v{c}}evi{\'c}, Milo{\v{s}} and Chiaro, Graziano and Cutini, Sara and Tosti, Gino",
    title = "{Classification of blazar candidates of uncertain type from the Fermi LAT 8-yr source catalogue with an artificial neural network}",
    eprint = "2002.10256",
    archivePrefix = "arXiv",
    primaryClass = "astro-ph.HE",
    doi = "10.1093/mnras/staa394",
    journal = "Mon. Not. Roy. Astron. Soc.",
    volume = "493",
    number = "2",
    pages = "1926--1935",
    year = "2020"
}

@ARTICLE{2019ApJ...887...18K,
       author = {{Kaur}, Amanpreet and {Falcone}, Abraham D. and {Stroh}, Michael D. and {Kennea}, Jamie A. and {Ferrara}, Elizabeth C.},
        title = "{Classification of New X-Ray Counterparts for Fermi Unassociated Gamma-Ray Sources Using the Swift X-Ray Telescope}",
      journal = {\apj},
         year = 2019,
        month = dec,
       volume = {887},
       number = {1},
          eid = {18},
        pages = {18},
          doi = {10.3847/1538-4357/ab4ceb},
archivePrefix = {arXiv},
       eprint = {1910.06317},
 primaryClass = {astro-ph.HE},
       adsurl = {https://ui.adsabs.harvard.edu/abs/2019ApJ...887...18K}
}

@ARTICLE{2020ApJS..248...23D,
       author = {{de Menezes}, Raniere and {D'Abrusco}, Raffaele and {Massaro}, Francesco and {Gasparrini}, Dario and {Nemmen}, Rodrigo},
        title = "{On the Physical Association of Fermi-LAT Blazars with Their Low-energy Counterparts}",
      journal = {\apjs},
         year = 2020,
        month = jun,
       volume = {248},
       number = {2},
          eid = {23},
        pages = {23},
          doi = {10.3847/1538-4365/ab8c4e},
archivePrefix = {arXiv},
       eprint = {2004.11236},
 primaryClass = {astro-ph.HE},
       adsurl = {https://ui.adsabs.harvard.edu/abs/2020ApJS..248...23D}
}

@article{Butter:2021mwl,
    author = {Butter, Anja and Finke, Thorben and Keil, Felicitas and Kr{\"a}mer, Michael and Manconi, Silvia},
    title = "{Classification of Fermi-LAT blazars with Bayesian neural networks}",
    eprint = "2112.01403",
    archivePrefix = "arXiv",
    primaryClass = "astro-ph.HE",
    reportNumber = "TTK-21-51",
    doi = "10.1088/1475-7516/2022/04/023",
    journal = "JCAP",
    volume = "04",
    number = "04",
    pages = "023",
    year = "2022"
}

@article{Agarwal:2023vra,
    author = "Agarwal, Aditi",
    title = "{Classification of Blazar Candidates of Unknown Type in Fermi 4LAC by Unanimous Voting from Multiple Machine-learning Algorithms}",
    eprint = "2303.14137",
    archivePrefix = "arXiv",
    primaryClass = "astro-ph.HE",
    doi = "10.3847/1538-4357/acbdfa",
    journal = "Astrophys. J.",
    volume = "946",
    number = "2",
    pages = "109",
    year = "2023"
}

@article{Bhatta:2023qrm,
    author = "Bhatta, Gopal and Gharat, Sarvesh and Borthakur, Abhimanyu and Kumar, Aman",
    title = "{Gamma-ray blazar classification using machine learning with advanced weight initialization and self-supervised learning techniques}",
    eprint = "2310.06095",
    archivePrefix = "arXiv",
    primaryClass = "astro-ph.HE",
    doi = "10.1093/mnras/stae028",
    journal = "Mon. Not. Roy. Astron. Soc.",
    volume = "528",
    number = "1",
    pages = "976--986",
    year = "2024"
}

@article{Gharat:2024uof,
  author       = {Gharat, Sarvesh and Borthakur, Abhimanyu and Bhatta, Gopal},
  title        = {Gamma Ray AGNs: Estimating Redshifts and Blazar Classification using traditional Neural Networks with smart initialization and self-supervised learning},
  eprint       = {2406.03782},
  archivePrefix= {arXiv},
  primaryClass = {astro-ph.HE},
  month        = {6},
  year         = {2024}
}

@article{Ballet:2023qzs,
    author = "Ballet, J. and Bruel, P. and Burnett, T. H. and Lott, B.",
    collaboration = "Fermi-LAT",
    title = "{Fermi Large Area Telescope Fourth Source Catalog Data Release 4 (4FGL-DR4)}",
    eprint = "2307.12546",
    archivePrefix = "arXiv",
    primaryClass = "astro-ph.HE",
    month = "7",
    year = "2023"
}

@article{Massaro:2015nia,
    author = "Massaro, Enrico and Maselli, Alessandro and Leto, Cristina and Marchegiani, Paolo and Perri, Matteo and Giommi, Paolo and Piranomonte, Silvia",
    title = "{The 5th edition of the Roma-BZCAT. A short presentation}",
    eprint = "1502.07755",
    archivePrefix = "arXiv",
    primaryClass = "astro-ph.HE",
    doi = "10.1007/s10509-015-2254-2",
    journal = "Astrophys. Space Sci.",
    volume = "357",
    number = "1",
    pages = "75",
    year = "2015"
}

@article{Fermi-LAT:2015bdd,
    author = "Ackermann, M. and others",
    collaboration = "Fermi-LAT",
    title = "{The Third Catalog of Active Galactic Nuclei Detected by the Fermi Large Area Telescope}",
    eprint = "1501.06054",
    archivePrefix = "arXiv",
    primaryClass = "astro-ph.HE",
    doi = "10.1088/0004-637X/810/1/14",
    journal = "Astrophys. J.",
    volume = "810",
    number = "1",
    pages = "14",
    year = "2015"
}

@article{Taylor_2017,
   title={TOPCAT: Desktop Exploration of Tabular Data for Astronomy and Beyond},
   volume={4},
   ISSN={2227-9709},
   url={http://dx.doi.org/10.3390/informatics4030018},
   DOI={10.3390/informatics4030018},
   number={3},
   journal={Informatics},
   publisher={MDPI AG},
   author={Taylor, Mark},
   year={2017},
   month=jun, pages={18} }

@misc{mao2023crossentropylossfunctionstheoretical,
      title={Cross-Entropy Loss Functions: Theoretical Analysis and Applications}, 
      author={Anqi Mao and Mehryar Mohri and Yutao Zhong},
      year={2023},
      eprint={2304.07288},
      archivePrefix={arXiv},
      primaryClass={cs.LG},
      url={https://arxiv.org/abs/2304.07288}, 
}

@misc{kingma2017adammethodstochasticoptimization,
      title={Adam: A Method for Stochastic Optimization}, 
      author={Diederik P. Kingma and Jimmy Ba},
      year={2017},
      eprint={1412.6980},
      archivePrefix={arXiv},
      primaryClass={cs.LG},
      url={https://arxiv.org/abs/1412.6980}, 
}

@misc{gorriz2024kfoldcrossvalidationbest,
      title={Is K-fold cross validation the best model selection method for Machine Learning?}, 
      author={Juan M Gorriz and R. Martin Clemente and F Segovia and J Ramirez and A Ortiz and J. Suckling},
      year={2024},
      eprint={2401.16407},
      archivePrefix={arXiv},
      primaryClass={stat.ML},
      url={https://arxiv.org/abs/2401.16407}, 
}

@ARTICLE{2020ApJ...892..105A,
       author = {{Ajello}, M. and {Angioni}, R. and {Axelsson}, M. and {Ballet}, J. and {Barbiellini}, G. and {Bastieri}, D. and {Becerra Gonzalez}, J. and {Bellazzini}, R. and {Bissaldi}, E. and {Bloom}, E.~D. and {Bonino}, R. and {Bottacini}, E. and {Bruel}, P. and {Buson}, S. and {Cafardo}, F. and {Cameron}, R.~A. and {Cavazzuti}, E. and {Chen}, S. and {Cheung}, C.~C. and {Ciprini}, S. and {Costantin}, D. and {Cutini}, S. and {D'Ammando}, F. and {de la Torre Luque}, P. and {de Menezes}, R. and {de Palma}, F. and {Desai}, A. and {Di Lalla}, N. and {Di Venere}, L. and {Dom{\'\i}nguez}, A. and {Dirirsa}, F. Fana and {Ferrara}, E.~C. and {Finke}, J. and {Franckowiak}, A. and {Fukazawa}, Y. and {Funk}, S. and {Fusco}, P. and {Gargano}, F. and {Garrappa}, S. and {Gasparrini}, D. and {Giglietto}, N. and {Giordano}, F. and {Giroletti}, M. and {Green}, D. and {Grenier}, I.~A. and {Guiriec}, S. and {Harita}, S. and {Hays}, E. and {Horan}, D. and {Itoh}, R. and {J{\'o}hannesson}, G. and {Kovac'evic'}, M. and {Krauss}, F. and {Kreter}, M. and {Kuss}, M. and {Larsson}, S. and {Leto}, C. and {Li}, J. and {Liodakis}, I. and {Longo}, F. and {Loparco}, F. and {Lott}, B. and {Lovellette}, M.~N. and {Lubrano}, P. and {Madejski}, G.~M. and {Maldera}, S. and {Manfreda}, A. and {Mart{\'\i}-Devesa}, G. and {Massaro}, F. and {Mazziotta}, M.~N. and {Mereu}, I. and {Meyer}, M. and {Migliori}, G. and {Mirabal}, N. and {Mizuno}, T. and {Monzani}, M.~E. and {Morselli}, A. and {Moskalenko}, I.~V. and {Negro}, M. and {Nemmen}, R. and {Nuss}, E. and {Ojha}, L.~S. and {Ojha}, R. and {Omodei}, N. and {Orienti}, M. and {Orlando}, E. and {Ormes}, J.~F. and {Paliya}, V.~S. and {Pei}, Z. and {Pe{\~n}a-Herazo}, H. and {Persic}, M. and {Pesce-Rollins}, M. and {Petrov}, L. and {Piron}, F. and {Poon}, H. and {Principe}, G. and {Rain{\`o}}, S. and {Rando}, R. and {Rani}, B. and {Razzano}, M. and {Razzaque}, S. and {Reimer}, A. and {Reimer}, O. and {Schinzel}, F.~K. and {Serini}, D. and {Sgr{\`o}}, C. and {Siskind}, E.~J. and {Spandre}, G. and {Spinelli}, P. and {Suson}, D.~J. and {Tachibana}, Y. and {Thompson}, D.~J. and {Torres}, D.~F. and {Torresi}, E. and {Troja}, E. and {Valverde}, J. and {van Zyl}, P. and {Yassine}, M.},
        title = "{The Fourth Catalog of Active Galactic Nuclei Detected by the Fermi Large Area Telescope}",
      journal = {\apj},
         year = 2020,
        month = apr,
       volume = {892},
       number = {2},
          eid = {105},
        pages = {105},
          doi = {10.3847/1538-4357/ab791e},
archivePrefix = {arXiv},
       eprint = {1905.10771},
 primaryClass = {astro-ph.HE},
       adsurl = {https://ui.adsabs.harvard.edu/abs/2020ApJ...892..105A}
}

@book{CoverThomas,
  author    = {Thomas M. Cover and Joy A. Thomas},
  title     = {Elements of Information Theory},
  publisher = {Wiley},
  edition   = {2nd},
  year      = {2006}
}
\bibliographystyle{aasjournalv7}

\end{document}